\documentclass[12pt]{article}
\usepackage{latexsym}
\usepackage{amsmath,amsbsy,amssymb}
\usepackage{verbatim}
\usepackage{bm}
\usepackage{cite}

\usepackage{curves}	
\usepackage{epic}
\usepackage{eepic}
\usepackage{epsfig}

\newcommand{\dd}{\mbox{\rm d}}

\newcommand{\dg}{\dagger}

\newcommand{\tl}{\tilde}

\newcommand{\p}{\partial}

\newcommand{\be}{\begin{equation}}
\newcommand{\bear}{\begin{eqnarray}}
\newcommand{\ear}{\end{eqnarray}}
\newcommand{\ee}{\end{equation}}

\newcommand{\lbl}{\label}
\newcommand{\bi}{\bibitem}
\newcommand{\ci}{\cite}

\newcommand{\vs}{\vspace}
\newcommand{\hs}{\hspace}

\newcommand{\vphi}{\varphi}

\newcommand{\vac}{|0 \rangle}

\newcommand{\om}{\omega}

\newcommand{\bp}{{\bm p}}
\newcommand{\bk}{{\bm k}}
\newcommand{\bx}{{\bm x}}
\newcommand{\hbp}{{\hat{\bm p}}}
\newcommand{\hbx}{{\hat{\bm x}}}
\newcommand{\tla}{{\tl a}}
\newcommand{\tlg}{{\tl g}}
\newcommand{\tlf}{{\tl f}}

\begin{document}

\begin{center}

\

\vs{.8cm}

\baselineskip .7cm

{\bf \Large Localized States in Quantum Field Theory} 

\vs{4mm}

\baselineskip .5cm
Matej Pav\v si\v c

Jo\v zef Stefan Institute, Jamova 39,
1000 Ljubljana, Slovenia

e-mail: matej.pavsic@ijs.si

\vs{3mm}
{\bf Abstract}
\end{center}

\baselineskip .43cm
{\footnotesize

Localized states in relativistic quantum field theories are usually considered as
problematic, because of their seemingly strange (non covariant) behavior
under Lorentz transformations, and because they can spread faster than light.
We point out that a careful quantum field theoretic analysis in which we
distinguish between basis position states and wave packet states
 clarifies
the issue of Lorentz covariance. The issue of causality is resolved by observing
that superluminal transmission of information cannot be achieved by such wave
packets. Within this context it follows that the Reef-Schlieder theorem,  which proves that
localized states can exhibit influence on each other over space like distances, does not imply
that such states cannot exist in quantum field theory.

\vs{2mm}

}

\baselineskip .55cm

\section{Introduction}

In non relativistic quantum mechanics and quantum field theories, states can be represented by
wave functions in configuration space. In the case of one particle states,  wave function is the
probability amplitude of finding the particle at a position $\bx$. In the literature there has been a
debate whether the analogous is possible in relativistic quantum
mechanics\,\ci{NewtonWigner,Wightman,Kalnay, Ruijgrok,Barat,Mir-Kasimov,Cirilo-Lombardo,Herrmann,Fleming1,Fleming2,Monahan}
and quantum field theory\,\ci{WightmanSchweber,Manoukian,Teller,Buscemi}.
Initially, when extending quantum mechanics to incorporate relativity, the subject of investigation was
the wave function satisfying the Klein-Gordon equation. Three main difficulties were encountered:

\ \ (i) The probability density was found to be either positive or negative.

\ (ii) Position operator\,\ci{NewtonWigner,Wightman,Kalnay, Ruijgrok,Barat,Mir-Kasimov,Cirilo-Lombardo,Herrmann,Monahan}
contained an extra term, which spoiled Lorentz covariance of such an operator.

(iii) Relativistic wave packets can spread faster than light, which has been interpreted as
violation of causality\,\ci{Hegerfeldt,Hegerfeldt2,Rosenstein,Mosley,Wagner,Eckstein,Barat,Buscemi,Ruijsenaars}.

With the advent of second quantization, the difficulty (i) was resolved  within the framework
of quantum field theory (QFT), in which instead of a wave function satisfying the Klein-Gordon equation,
one has an operator-valued non Hermitian field that creates particles and antiparticles of opposite 
electric charge.

It is usually believed that  in QFT states cannot
be localized, so that QFT solves the difficulties (ii) and (iii) as well. But such a claim has
to be confronted with the fact that relativistic quantum mechanics for positive frequencies is embedded in quantum field theory\,\ci{Al-Hashimi,Gavrilov}.
A consequence is that the localized states, either as wave packets picked around a certain
spatial region, or exactly confined within it, also occur in quantum field theory. We will show why such states
are not problematic at all.  As is well known\,\ci{Peskin},
the Fock space states with definite momentum, $a^\dg (\bp) \vac$, $a^\dg (\bp_1) a^\dg (\bp_2)...\vac$,
can be superposed by means of complex valued  wave packet profiles $g(t,\bp)$, $g(t,\bp_1,\bp_2)$,...,
into the states with indefinite momentum. We choose to put time dependence on $g$, and leave $a^\dg$
independent of time.  Even if we consider the case of a Hermitian scalar field $\vphi (x) = \vphi^\dg (x)$
(i.e., if the classical field that we quantize is real), the wave packet profile is in general complex
and satisfies $g^* g >0$. Similarly, in the case of harmonic oscillator, we quantize the classical {\it real}
variable $x(t)$, whilst a generic quantum state is a superposition of the states $a^\dg \vac$,
$a^\dg a^\dg \vac,...,(a^\dg)^n \vac$, the superposition coefficients $c_n$ being {\it complex} and
satisfying $c_n^* c_n >0$.
The problem of the first quantization in which the probability density can be negative, does not
exist in the second quantized, i.e., quantum field theory. What is negative in QFT, is the
zero component, $j^0$, of the current  density $j^\mu$, interpreted as a charge current density,
whilst the probability density, which is given
in terms of a wave packet profile, is always positive\,\ci{Horwitz,Fonda}. This is a straightforward consequence of the fact
that the Hamilton operator is positive definite with respect to the Fock space states, created by
the action of $a^\dg (\bp)$ on the vacuum, which is annihilated by $a(\bp)$. Therefore, only positive frequency
wave functions occur in quantum field theory, a fact which
is often overlooked in the literature.

In an appropriate normalization\,\ci{Horwitz,Fonda}, a wave packet profile $g(t,\bp)$ gives the scalar product
$\int \dd^3 \bp \, g^*(t,\bp) g(t,\bp)$. It can be Fourier transformed into a position space
wave packet profile $f(t,\bx)$, giving  $\int \dd^3 \bx f^*(t,\bx) f(t,\bx)$. The absolute square
$|f(t,\bx)|^2= f^* (t,\bx) f(t,\bx)$  gives
the probability density of finding the particles at the position $\bx$. Similarly, the corresponding momentum space
creation and annihilation operators, $a^\dg (\bp)$, $a (\bp)$, can be Fourier transformed
into the position space operators\footnote{
For simplicity we use here the same symbol $a$ for the Fourier transformed operators as well.}
 $a^\dg (\bx)$, $a^\dg (\bx)$, with $a^\dg (\bx) \vac$ being a
state with the particle at position $\bx$. The wave packet profile at, say $t=0$, is then
$f(0,\bx') = \delta^3 (\bx' - \bx)$.

In the absence of interaction it makes sense to consider single particle states only. As any multiparticle state,
also a single particle state $\int \dd^3 \bx \, f(t,\bx) a^\dg (\bx) \vac$
satisfies the Schr\"odinger equation with the Hamilton operator of the considered quantum field
theory (i.g., that of a scalar field), and it turns out, as already pointed out in Ref.\,\ci{Al-Hashimi,Gavrilov},
that $f(t,\bx)$ satisfies the Klein-Gordon equation for positive frequencies. Relativistic quantum
mechanics for positive frequencies (energies) is thus automatically embedded in QFT, which therefore
inherits all the issues concerning state localization and causality.
As pointed out by Valente\,\ci{Valente},
there are several distinct concepts of causality in the literature, and not all of them imply faster than light
transmission of {\it information} which only can lead to causality paradoxes. The `causality'
used in algebraic (axiomatic) quantum field theory\,\ci{Haag} as one of the axioms is of such a kind
that its violation is not problematic. Consequently, the Reeh-Schlieder\,\ci{Reeh-Schlieder}
theorem does not violate relativistic causality\,\ci{Valente} and hence does not imply
that states cannot be localized in a finite region.

The  initially  $\delta$-function wave packet evolves with time as the relativistic Green function
$G(t,x;0,0)$ considered in Refs.\,\ci{Fonda,Al-Hashimi}. This has
to be taken into account when transforming $f(0,\bx)$ into another Lorentz
frame. It comes out that if at the initial time $t=0$ in a Lorentz frame $S$ a particle is localized at $\bx=0$, 
then from the perspective of another Lorentz frame $S'$ the same particle is also
localized in the same spacetime point. In the case
of a boost, the frame $S'$ is merely pseudo rotated with respect to $S$, so that both reference frames have the same
origin, and thus in $S'$ the particle at $t'=0$ is localized at $\bx'=0$. This is a consequence of the
properties of $G(t,x;0,0) \equiv G(t,x)$ whose absolute square $|G(t,x)|^2$ is singular on the light cone,
and zero everywhere else. Initially, the particle is thus localized in the ``origin'' of the light cone, regardless
of the Lorentz reference that it is observed from. At latter time, the particle is localized on the
intersection of a hypersurface $\Sigma$ with the light cone. If  $\Sigma$
is a simultaneity hypersurface in the frame $S$, it is no longer a simultaneity hypersurface in the frame
$S'$. Therefore,  an observer in $S'$ must consider the wave packet on his simultaneity hypersurface
$\Sigma'$, in order to see how the wave packet is localized in $S'$.

If a wave packet is not $\delta$-like, but spread, then its behavior\,\ci{Fonda,Rosenstein,Al-Hashimi}
depends on whether its width is smaller or greater than the Compton wavelength.
If it is smaller, then its probability density after some time becomes concentrated in
the vicinity of the light cone, and not exactly on the light cone as in the case of zero width.
If the wave packet width is greater than the Compton wave length, then the wave packet's
probability density is concentrated around the particle's classical world line. The particle
is localized (in the sense of being peaked) on the intersection of a hypersurface with the spacetime distribution of the
probability density. The choice of the hypersurface depends on the Lorentz frame in
which we observe the wave packet evolution. Nothing unusual happens if we go into
another Lorentz frame: the particle wave function is still a wave packet spread around
the classical trajectory, which in a different Lorentz frame has a different velocity. It is important to stress that
we use the term ``localization'' in a broad sense, either as (i) point-like localization, (ii) localization
in a finite spatial region and vanishing outside, (iii) localization in a finite region
decaying outside, and (iv) ``effective'' localization like a Gaussian wave packet. We will show that
none of those kinds of localization is problematic

In this paper we consider wave packet profiles in the free scalar field theory, revise
the role of position operator and the behavior of states under Lorentz transformations.
We find that localized states are not problematic at all. We
also calculate some explicit examples of wave packets for the widths greater and smaller than
the Compton wave length. Finally we discuss the issue of causality violation in the cases when
the probability density leaks outside the light cone. We argue that in order to violate causality
one has to be able to transmit information faster than light, and that this cannot be achieved by
means of the wave packets whose centroid position is on, or inside, the light cone.
To transmit information, one wave packet is not enough; it is necessary to have a modulated beam
of particles, which can be achieved by sending one wave packet after another. Because their centers
move  with the velocity of light or slower, a train of the wave packets which bears a message, cannot
travel faster than light. Some other authors\,\ci{Fleming2,Wagner} also had similar ideas.
However, there could exist other ingenious ways to use relativistic wave
packets to send signals faster than light. But since the Compton length is of a subatomic size,
such signals could be sent into a very nearby past only, so that no macroscopic effects of the grand father
causality paradox could take place.

\section{Wave packet profiles in the free scalar field theory}

To make the paper self-consistent and to clarify certain confusion regarding state
localization, we will review the essential features of the free scalar field theory.
Let us consider a real scalar field $\vphi$,  $x \equiv (t,\bx)$, described by the action
\be
   I[\vphi(x)] = \frac{1}{2} \int \dd^4 x\, (\p_\mu \vphi \p^\mu \vphi - m^2 \vphi^2 ).
\lbl{2.1}
\ee
Variation of the latter action with respect to $\vphi(x)$ gives the Klein-Gordon equation
\be
  \p_\mu \p^\mu \vphi + m^2 \vphi = 0.
\lbl{2.2}
\ee

From the canonically conjugated variables $\vphi(t,\bx)$, $\Pi(t,\bx) = \p{\cal L}/\p {\dot \vphi} = 
{\dot \vphi}$, we can construct the Hamiltonian
\be
  H=  \int \dd^3 \bx (\Pi {\dot \vphi} - {\cal L}) =
  \frac{1}{2} \int \dd^3 \bx (\Pi^2 - \p^i \vphi \p_i \vphi + m^2 \vphi^2),
\lbl{2.3}
\ee
where $\p_i \equiv \p/\p x^i$, $i=1,2,3$.

Using the Poisson bracket relations
\be
  \lbrace \vphi (t,\bx),\Pi(t,\bx') \rbrace = \delta^3 (\bx - \bx'),
\lbl{2.4a}
\ee
\be
  \lbrace \vphi (t,\bx),\vphi(t,\bx') \rbrace = 0~,~~~~\lbrace \Pi (t,\bx),\Pi(t,\bx') \rbrace =0,
\lbl{2.4b}
\ee
we find that the equations of motion
\be
  {\dot \vphi}=  \lbrace \vphi, H \rbrace ~,~~~~~{\dot \Pi} =  \lbrace \Pi , H \rbrace ,
\lbl{2.5}
\ee
are equivalent to the Klein-Gordon equation (\ref{2.2}).

A general solution of the Klein-Gordon equation is
\be
   \vphi (x) = \int \frac{\dd^4 k}{(2 \pi)^4} c(k) \delta (k^2 - m^2) {\rm e}^{- i kx} 
  = \int \frac{\dd^3 \bk}{(2 \pi)^3 2 \om_\bk } \left ( {\tl a}(\bk) {\rm e}^{-i kx} + 
  {\tl a}^* (\bk) {\rm e}^{i kx} \right ),
\lbl{2.5a}
\ee
where $\om_\bk = \sqrt{m^2 + \bk^2} > 0$, and
\be
  {\tl a} (\bk) = \frac{1}{2 \pi} c (\om_\bk, \bk)~,~~~~{\tl a}^* (\bk) = \frac{1}{2 \pi} c (-\om_\bk, \bk) .
\lbl{2.5b}
\ee

Upon quantization, $\vphi$ and $\Pi$  become operators satisfying
\be
  [ \vphi (t,\bx),\Pi(t,\bx') ] = i \delta^3 (\bx - \bx'),
\lbl{2.6}
\ee
\be
  [ \vphi (t,\bx),\vphi(t,\bx') ] = 0~,~~~~ [ \Pi (t,\bx),\Pi(t,\bx') ] =0 .
\lbl{2.7}
\ee
The Klein-Gordon equation (\ref{2.2}) is now the equation of motion for  the operator $\vphi (x)$,
and is equivalent to  the Heisenberg equations of motion
\be
  {\dot \vphi} = -i  [\vphi, H] ~,~~~~~{\dot \Pi} =-i [\Pi,H]
\lbl{2.8}
\ee
that are quantum analog of the classical equations (\ref{2.5}).

The quantum field $\vphi (x)$ that solves the Klein-Gordon equation can be expanded according to
\be
   \vphi (x) =  \int \frac{\dd^3 \bk}{(2 \pi)^3 2 \om_\bk } \left ( {\tl a}(\bk) {\rm e}^{-i kx} + 
  {\tl a}^\dg (\bk) {\rm e}^{i kx} \right ),
\lbl{2.9}
\ee
where ${\tl a} (\bk)$ and ${\tl a}^\dg (\bk)$ are operators satisfying
\be
  [ {\tl a}(\bk), {\tl a}^\dg (\bk')] = (2 \pi)^3 2 \om_\bk \delta^3 (\bk - \bk'),
\lbl{2.10}
\ee
\be
  [ {\tl a}(\bk), {\tl a} (\bk')] = 0~,~~~~~~~ [ {\tl a}^\dg (\bk), {\tl a}^\dg (\bk')] = 0.
\lbl{2.11}
\ee
The latter commutation relations for ${\tl a}(\bk)$, $\tla^\dg (\bk)$ are consistent with the
commutation relations (\ref{2.6}),(\ref{2.7}) for $\vphi (x)$, $\Pi (x)$.

The Hamilton operator, given by the expression (\ref{2.3}), can be rewritten in terms
of $\tla (\bk)$, $\tla^\dg (\bk)$:
\bear
 && H = \frac{1}{2} \int \frac{\dd^3 \bk}{(2 \pi)^3 2 \om_\bk } \om_\bk \left ( \tla^\dg (\bk) \tla (\bk)
  + \tla (\bk) \tla^\dg (\bk) \right ) \nonumber\\
   && \hs{6mm} = \int \frac{\dd^3 \bk}{(2 \pi)^3 2 \om_\bk } \om_\bk \left ( \tla^\dg (\bk) \tla (\bk) +
   \frac{\delta (0)}{2} \right ) .
\lbl{2.12}
\ear

If we define vacuum according to
\be
  \tla (\bk) \vac = 0 ,
\lbl{2.13}
\ee
then the vacuum expectation value of the Hamiltonian is
\be
  \langle 0| H \vac = \frac{1}{2} \int \frac{\dd^3 \bk}{(2 \pi)^3 2 \om_\bk } \om_\bk \delta (0)
  = E_0 = \infty .
\lbl{2.14}
\ee

A generic state is a superposition of the basis states created by $\tla^\dg (\bk)$:
\be
  |\Psi \rangle = \sum_n \int \frac{\dd^3 \bk_1 \dd^3 \bk_2...\dd^3 \bk_n }{(2 \pi)^{3 n} 2 \om_{\bk_1}...2 \om_{\bk_n} } \tlg(t,\bk_1,\bk_2,...,\bk_n)
  \tla^\dg (\bk_1) \tla^\dg (\bk_2) ...\tla^\dg (\bk_n) \vac ,
\lbl{2.15}
\ee
where $\tlg (t,\bk_1,\bk_2,...,\bk_n)$ is a complex valued wave packet profile.

A single particle state is
\be
  |\Psi \rangle = \int \frac{\dd^3 \bk}{(2 \pi)^3 2 \om_\bk } \tlg(t,\bk)
  \tla^\dg (\bk)  \vac .
\lbl{2.16}
\ee
It evolves according to the Schr\"odinger equation
\be
  i \frac{\p |\Psi \rangle}{\p t} = H |\Psi \rangle ,
\lbl{2.17}
\ee
where the Hamilton operator is given in Eq.\,(\ref{2.12}).  From the latter equation, by using (\ref{2.13})
and the commutation relations (\ref{2.10}),(\ref{2.11}), we obtain the following equation
of motion for the wave packet profile\,\ci{Schweber}, p.\,162, \ci{Rosenstein}
\be
  i \frac{\p \tlg (t,\bk)}{\p t} = (\om_\bk + E_0) \tlg(t,\bk),
\lbl{2.18}
\ee
whose solution is
\be
  \tlg(t,\bk) = {\rm e}^{- i (\om_\bk + E_0)t} \tlg (\bk) .
\lbl{2.19}
\ee

The scalar product of a single particle state is
\be
  \langle \Psi|\Psi \rangle = \int \frac{\dd^3 \bk}{(2 \pi)^3 2 \om_\bk } \tlg^*(t,\bk) \tlg (t,\bk),
\lbl{2.20}
\ee
where the zero point energy, $E_0$, cancels out. Therefore, from now on we will omit $E_0$ in
the expression (\ref{2.18}), and assume
\be
  i \frac{\p \tlg (t,\bk}{\p t} = \om_\bk \tlg(t,\bk),
\lbl{2.19b}
\ee
\be
  \tlg(t,\bk) = {\rm e}^{- i \om_\bk t} \tlg (\bk) .
\lbl{2.19a}
\ee
Let us now project a single particle state $|\Psi \rangle$ (Eq.\,(\ref{2.16})) onto a basis
state $|{\tl \bk} \rangle$, defined according to
\be
  |{\tl \bk} \rangle = \tla^\dg (\bk) \vac ~,~~~\langle {\tl \bk}| = \langle 0|\tla (\bk) .
\lbl{2.21}
\ee
We obtain
\be
  \langle {\tl \bk}|\Psi \rangle = \langle 0|\tla (\bk) \int \frac{\dd^3 \bk'}{(2 \pi)^3 2 \om_\bk' } \tlg(t,\bk')
  \tla^\dg (\bk')  \vac = \tlg(t,\bk),
\lbl{2.22}
\ee
where we have taken into account the commutation relation (\ref{2.10}) and the vacuum
property (\ref{2.13}).

We can also project $|\Psi \rangle$ onto a state $|{\tl \bx} \rangle$ defined according to
\be
  |{\tl \bx} \rangle = \vphi(0,\bx) \vac = \vphi^{(+)}(0,\bx) \vac \equiv \tla^\dg (\bx) \vac ,
\lbl{2.23}
\ee
\be
  \langle {\tl \bx}| = \langle 0|\vphi(0,\bx) = \langle 0| \vphi^{(-)} (0,\bx) \equiv \langle 0|\tla (\bx) .
\lbl{2.24}
\ee
Here $\vphi^{(+)}$ and $\vphi^{(-)}$ are, respectively, the positive and negative frequency part
of $\vphi(x)$, given in Eq.\,(\ref{2.9}). We then have
\be
  \langle {\tl \bx}|\Psi \rangle = \langle 0|\vphi^{(-)}(0,\bx)|\Psi \rangle 
  =\int \frac{\dd^3 \bk}{(2 \pi)^3 2 \om_\bk } {\rm e}^{ -i \bk \bx} \tlg (t,\bk) = \tlf (t,\bx),
\lbl{2.25}
\ee
and
\be
  |\Psi \rangle = \int |{\tl \bk} \rangle \frac{\dd^3 \bk}{(2 \pi)^3 2 \om_\bk }  \langle {\tl \bk}|\Psi \rangle
  =\int |{\tl \bx}  \rangle \dd^3 \bx \,2 \sqrt{m^2- \nabla^2}\,\langle {\tl \bx}|\Psi \rangle .
\lbl{2.26}
\ee

In Eq.\,(\ref{2.25}) we have the transformations from the amplitude $\tlg (t,\bk)$ to
$\tlf (t,\bx)$. The inverse transformation is
\be
  \tlg (t,\bk) = 2 \om_\bk \int \dd^3 \bx \, {\rm e}^{i \bk \bx} \tlf (t,\bx) .
\lbl{2.27}
\ee
If we insert the latter expression into the scalar product (\ref{2.20}), we obtain
\bear
  &&\langle \Psi|\Psi \rangle = \int \dd^3 \bk \,\dd^3 \bx \, \dd^3 \bx'\, 2 \om_\bk \frac{{\rm e}^{i \bk (\bx-\bx')}}{(2 \pi)^3}
  \tlf^* (t,\bx) \tlf (t,\bx')  \nonumber \\
  &&\hs{1cm}=\int \dd^3 \bx \, \dd^3 \bx'\, 2 \sqrt{m^2 + (-i \nabla)^2} \, \delta^3 (\bx-\bx')
   \tlf^* (t,\bx) \tlf (t,\bx')  \nonumber \\
  &&\hs{1cm}=\int \dd^3 \bx \, \dd^3 \bx'\, \delta^3 (\bx-\bx') 
   \tlf^* (t,\bx) 2 \sqrt{m^2 + (-i \nabla)^2} \,\tlf (t,\bx')  \nonumber \\
   &&\hs{1cm}= \int \dd^3 \bx\, \tlf^* (t,\bx) 2 \sqrt{m^2 + (-i \nabla)^2} \,\tlf (t,\bx') \nonumber\\
  &&\hs{1cm}= \int \dd^3 \bx\, \left [ \tlf^* (t,\bx) \sqrt{m^2 + (-i \nabla)^2} \,\tlf (t,\bx) \right . \nonumber\\
   &&\hs{3cm}+ \left . \left ( \sqrt{m^2 + (-i \nabla)^2} \,\tlf^* (t,\bx) \right ) \tlf(t,\bx) \right ] .
\lbl{2.28}
\ear

Let us now use Eq.\,(\ref{2.19b}), from which we obtain
\be
  i \frac{\p}{\p t} \int \frac{\dd^3 \bk}{(2 \pi)^3 2 \om_\bk } {\rm e}^{-i \bk \bx }\tlg (t,\bk)
  =\int \frac{\dd^3 \bk}{(2 \pi)^3 2 \om_\bk } \,\om_\bk \,\tlg (t,\bk) {\rm e}^{-i \bk \bx} .
\lbl{2.29}
\ee
Using (\ref{2.25}), Eq.\,(\ref{2.29}) gives the well known relativistic Schr\"odinger
equation\,\ci{Mosley},\ci{Fonda,Horwitz}
\be
   i \frac{\p}{\p t} \tlf (t,\bx) = \sqrt{m^2 +(\-i \nabla)^2} \, \tlf (t,\bx) .
\lbl{2.30}
\ee
The scalar product is thus
\be
  2 i  \int \dd^3 \bx \, \tlf^* (t,\bx)  \frac{\p}{\p t} \tlf (t,\bx) =
    i \int \dd^3 \bx \left ( \tlf^* (t,\bx) \frac{\p}{\p t} \tlf (t,\bx) - \frac{\p}{\p t} \tlf^* (t,\bx) \tlf (t,\bx) \right ).
\lbl{2.31}
\ee

We can do the calculation in the opposite way and start from Eq.\,(\ref{2.31}).
Inserting the expression  (\ref{2.25}) for $\tlf (t,\bx)$, and using Eq.\,(\ref{2.19b}),
we have
\bear
  &&2 i  \int \dd^3 \bx \, \tlf^* (t,\bx)  \p_0 \tlf (t,\bx) 
  = \int \dd^3 \bx \frac{\dd^3 \bk}{(2 \pi)^3 2 \om_\bk }  \,\tlg (t,\bk) {\rm e}^{i \bk \bx} 
   \frac{\dd^3 \bk'}{(2 \pi)^3 2 \om_\bk' }  \,\p_0 \tlg (t,\bk') {\rm e}^{-i \bk' \bx} \nonumber\\
  &&\hs{4.7cm} = \int \frac{\dd^3 \bk}{(2 \pi)^3 2 \om_\bk }  \,\tlg^* (t,\bk) \tlg (t,\bk)
\lbl{2.32}
\ear
Because the right hand side of the latter equation is Lorentz invariant, also the left
hand side is Lorentz invariant. This can be also seen if we rewrite the expression
(\ref{2.31}) in a covariant way as $2 i  \int \dd^3 \bx \, \tlf^*  \p_0 \tlf =
2 i \int \dd \Sigma^0 \, \tlf^*  \p_0 \tlf = 2 i \int \dd \Sigma^\mu \, \tlf^*  \p_\mu \tlf $,
where in this particular Lorentz frame it is $\dd \Sigma^\mu = (\dd \Sigma^0, 0,0,0)$.

The scalar product so defined is positive, because the wave packet profile
${\tl g}(t,\bk)$ satisfies the Schr\"odinger equation (\ref{2.18}) with positive energy.
This is so because the Hamilton operator (\ref{2.12}) is positive definite with
respect to the states created by ${\tl a}^\dg (\bk)$, and because the vacuum satisfies
${\tl a} (\bk) \vac = 0$. Analogous holds for a Fourier-like transformed wave packet
${\tl f} (t,\bx)$ and the operators ${\tl a}^\dg (\bx)$, $a (\bx)$, defined
in Eqs.\,(\ref{2.23})--(\ref{2.25}).

\section{An alternative normalization of the operators and wave packets}

If instead of $\tla (\bk)$  and $\tlg (t,\bk)$ we introduce\footnote{Such a normalization
is also used in the literature, e.g., in the textbook by Peskin\,\ci{Peskin}.}
\be
  a({\bk}) = \frac{\tla (\bk)}{\sqrt{(2 \pi)^3 2 \om_\bk}}~,
  ~~~~~~~g(t,\bk) = \frac{\tlg (\bk)}{\sqrt{(2 \pi)^3 2 \om_\bk}} ,
\lbl{3.1}
\ee
and analogous for $a^\dg (t,\bk)$, $g^* (t,\bk)$, then many expressions and derivations
become much simpler.

The field operator becomes
\be
   \vphi (x) =  \int \frac{\dd^3 \bk}{\sqrt{(2 \pi)^3 2 \om_\bk }} \left ( {a}(\bk) {\rm e}^{-i kx} + 
  {a}^\dg (\bk) {\rm e}^{i kx} \right ),
\lbl{3.2}
\ee
where $a(\bk)$ and $a^\dg (\bk)$ satisfy
\be
  [a(\bk),a^\dg (\bk')] = \delta (\bk-\bk') ,
\lbl{3.3}
\ee
\be
  [a(\bk),a (\bk')] =0 ~,~~~~~~[a^\dg(\bk),a^\dg (\bk')] =0 ,
\lbl{3.4}
\ee
so that the Hamiltonian is now
\be
  H = \frac{1}{2} \int \dd^3 \bk \, \om_\bk \,\left ( a^\dg (\bk) a (\bk) + a (\bk) a^\dg (\bk) \right ).
\lbl{3.5}
\ee

A generic single particle state (\ref{2.16}) can be rewritten as
\be
  |\Psi \rangle = \int \dd^3 \bk \, g(t,\bk) a^\dg (\bk) \vac .
\lbl{3.6}
\ee
From the Schr\"odinger equation (\ref{2.10}) it now follows that\,\ci{Rosenstein}
\be
  i \frac{\p g (t,\bk}{\p t} = \om_\bk g(t,\bk),
\lbl{3.7}
\ee
\be
  g(t,\bk) = {\rm e}^{-i \om_\bk t} g (\bk) .
\lbl{3.8}
\ee

The Fourier transformed quantities are
\be
  a(\bx) = \frac{1}{\sqrt{(2 \pi)^3}} \int \dd^3 \bk \, a (\bk) {\rm e}^{i \bk \bx} ,
\lbl{3.9}
\ee
\be
  f(t,\bx) = \frac{1}{\sqrt{(2 \pi)^3}} \int \dd^3 \bk \, f(t, (\bk) {\rm e}^{i \bk \bx} ,
\lbl{3.10}
\ee
where $f(t,\bx)$ satisfies
\be
  i \frac{\p f (t,\bx)}{\p t} = \sqrt{m^2 + (-i \nabla)^2} f(t,\bx) .
\lbl{3.10a}
\ee
Analogous expressions hold for $a^\dg (\bx)$ and $f^* (t,x)$.

Using (\ref{3.3}),(\ref{3.4}) and (\ref{3.9}), we find that $a(\bx)$ and $a^\dg (\bx)$ satisfy the
following commutation relations:
\be
  [a(\bx),a^\dg (\bx')] = \delta (\bx-\bx') ,
\lbl{3.9a}
\ee
\be
  [a(\bx),a (\bx')] =0 ~,~~~~~~[a^\dg(\bx),a^\dg (\bx')] =0 ,
\lbl{3.9b}
\ee

A generic single particle state (\ref{3.6}) can be re-expressed as a superposition of
the states created by $a^\dg (\bx)$:
\be
  |\Psi \rangle = \int \dd^3 \bx \, f(t,\bx) a^\dg (\bx) \vac .
\lbl{3.11}
\ee

The scalar product becomes
\be
  \langle \Psi|\Psi \rangle = \int \dd^3 \bk \, g^* (t,\bk) g (t,\bk) =
  \int \dd^3 \bx \, f^* (t,\bx) f(t,\bx) .
\lbl{3.12}
\ee
It is of course Lorentz invariant, though in the above form does not manifestly look so,
because $g(t,\bk)$ and $f(t,\bx)$ do not have simple Lorentz transformations\,\ci{Fonda,Horwitz}.
But the quantities $\tlg (t,\bk)$ and $\tlf (t,\bx) \equiv \tlf (x)$ are scalars:
\be
   \tlg'(t',\bk') = \tlg (t,\bk) ~. ~~~~~~~\tlf'(\Lambda x) = \tlf (x) ,
\lbl{3.13}
\ee
where $t',\bk'$ and $x' = \Lambda x$ are Lorentz transformed quantities.

The transformation between $g(t,\bk)$ and $\tlg(t,\bk)$ is simple, namely (\ref{3.1}),
whilst the transformation between $f(t,\bx)$ and $\tlf (t,\bx)$ is\,\ci{Fonda,Horwitz}
\be
  f(t,\bx) = \int \dd^3 \bx' \, K(\bx,\bx') \tlf (t,\bx') ,
\lbl{3.14}
\ee
where
\be
  K(\bx,\bx') = \int \frac{\dd^3 \bk \, \sqrt{2 \om_\bk} \,{\rm e}^{i \bk (\bx-\bx')}}{(2 \pi)^3} ,
\lbl{3.15}
\ee
which gives\,\ci{Fonda,Horwitz}
\be
  f(t,\bx) = \sqrt{2} \left ( m^2 + (-i \nabla)^2 \right )^{1/4} \tlf (t,\bx) .
\lbl{3.16}
\ee
The latter transformation can be straightforwardly derived from Eqs.\,(\ref{3.10}),
(\ref{2.27}) and (\ref{3.1}).

The inverse transformations is
\be
     \tlf(t,\bx) = \int \dd^3 \bx \, K^{-1}(\bx,\bx') f (t,\bx') =
     \frac{1}{\sqrt{2}} \left ( m^2 + (-i \nabla)^2 \right )^{-1/4}  f(t,\bx) ,
\lbl{3.16a}
\ee
with
\be
  K^{-1} (\bx,\bx') = \int \frac{\dd^3 \bk}{(2 \pi)^3} \frac{{\rm e}^{i \bk (\bx-\bx')}}{\sqrt{2 \om_\bk}}   ,
\lbl{3.16b}
\ee
satisfying
\be
  \int K(\bx,\bx'') \,\dd^3 \bx'' K^{-1} (\bx'',\bx') = \delta^3 (\bx - \bx') .
\lbl{3.16c}
\ee

Analogous transformation also holds for the creation/annihilation operators.
Denoting ${\tl a} (\bx) \equiv \vphi^- (0,\bx)$, ${\tl a}^\dg (\bx) \equiv \vphi^+ (0,\bx)$,
we have
\be
  a(\bx) = \int \dd^3 \bx' \, K(\bx,\bx') {\tl a} (\bx') 
  = \sqrt{2} \left ( m^2 + (-i \nabla)^2 \right )^{1/4} {\tl a} (\bx)
\lbl{3.14a}
\ee
\be
  a^\dg (\bx) = \int \dd^3 \bx' \, K(\bx,\bx') {\tl a}^\dg (\bx') 
  = \sqrt{2} \left ( m^2 + (-i \nabla)^2 \right )^{1/4} {\tl a}^\dg (\bx) ,
\lbl{3.14b}
\ee

If in Eq.\,(\ref{3.12}) we express $g(t,\bk)$ according to (\ref{3.1}) and $f(t,\bx)$ according
to (\ref{3.16}), we obtain the scalar product in the form (\ref{2.20}) or (\ref{2.31}), as
we should. To recapitulate, the single state scalar product can be expressed in the following
four ways:
\bear
  &&\langle \Psi |\Psi \rangle = \int \frac{\dd^3 \bk}{(2 \pi)^3 2 \om_\bk}\,\tlg^* (t,\bk) \tlg(t.\bk)
  = 2 i \int \dd^3 \bx \, \tl f^* (t,\bx) \p_0 \tlf(t,\bx)  \nonumber \\
  &&\hs{2cm}=\int \dd^3 \bk \, g^* (t,\bk) g(t,\bk) = \int \dd^3 \bx \, f^* (t,\bx) f(t,\bx) ,
\lbl{3.17}
\ear
where the state is given as
\bear
 && |\Psi \rangle = \int \frac{\dd^3 \bk}{(2 \pi)^3 2 \om_\bk}\, \tlg(t.\bk) \tla^\dg (\bk) \vac =
  \int \dd^3 \bx \, \tlf (t,\bx) 2 \sqrt{m^2 + (-i \nabla)^2} \,\vphi  (0,\bx) \vac \nonumber \\
  &&\hs{1cm}= \int \dd^3 \bk \, g (t,\bk) a^\dg (\bk) \vac = \int \dd^3 \bx \, f(t,\bx) a^\dg (\bx) \vac .
\lbl{3.18}
\ear
It  contains positive energy basis states only, negative ones are excluded, bacause $\tla (\bk) \vac = 0$.
The scalar product is positive.

The transformation between $\tlf(t,\bx)$ and $f(t,\bx)$ is non local. Thus, if $\tlf(t,x)$ at $t=0$
is a localized function of $\bx$,
\be
  \tlf (0,\bx) = \delta^3 (\bx-\bx_0) ,
\lbl{3.19}
\ee
then $f(0,\bx)$ is a delocalized function of $\bx$ according to
\be
 f(0,\bx) = \int \frac{\dd^3 \bk \, \sqrt{2 \om_\bk} \,{\rm e}^{i \bk (\bx-\bx_0)}}{(2 \pi)^3}
 = \sqrt{2} \left ( m^2 + (-i \nabla)^2 \right )^{1/4} \delta^3 (\bx-\bx_0) .
\lbl{3.20}
\ee
On the contrary, if $f(t,\bx)$ at $t=0$ is
\be
  f(0,\bx) = \delta^3 (\bx-\bx_0),
\lbl{3.20a}
\ee
then $\tlf(0,\bx)$ is a delocalized function of $\bx$:
\be
  \tlf(0,\bx) = \int \frac{\dd^3 \bk}{(2 \pi)^3} \frac{{\rm e}^{i \bk (\bx-\bx_0)}}{\sqrt{2 \om_\bk}}
  = \frac{1}{\sqrt{2}} \left ( m^2 + (-i \nabla)^2 \right )^{-1/4} \,\delta^3 (\bx - \bx_0) .
\lbl{3.21}
\ee
Despite being delocalized in $\bx$, the latter function is an eigenfunction of the position
operator\,\ci{NewtonWigner,Wightman,Kalnay, Ruijgrok,Barat,Mir-Kasimov,Cirilo-Lombardo,Herrmann},
and it represents a state, localized at position
$\bx_0$. But so does the function $f(0,\bx)$ of Eq.\,(\ref{3.20a}), which, as we will see,
is also an eigenstate of the position operator. We see that representation of a state
in terms of $f(t,\bx)$ is better adapted for description of a wave packet  state, effectively
localized within a finite spatial region.
In the next section we will discuss  properties of the position operator and localized
states in free scalar field theory.

\section{Position and momentum operator}

In previous sections we represented a generic single particle state as a superposition
(\ref{3.18}) of the basis states, created either by $a^\dg(\bx)$, $a^\dg (\bp)$, or,
$\tla^\dg (\bx)\equiv \vphi^{(+)}(0,\bx)$,$\tla^\dg (\bp)$:
\be
  |\bx \rangle = a^\dg (\bx) \vac ~,~~~~~~~|\bp \rangle = a^\dg (\bp) \vac ,
\lbl{4.1}
\ee
\be
  |{\tl \bx} \rangle = \tla^\dg (\bx) \vac ~,~~~~~~~
  |{\tl \bp} \rangle = \tla^\dg (\bp) \vac ,
\lbl{4.2}
\ee
the corresponding wave packet profiles being
\be
  f(t,\bx)= \langle \bx |\Psi \rangle ~,~~~~~~~g(t,\bp) = \langle \bp |\psi \rangle ,
\lbl{4.3}
\ee
\be
 \tlf(t,\bx) = \langle {\tl \bx}|\Psi \rangle ~,~~~~~~~\tlg(t,\bp) = \langle {\tl \bp}|\psi \rangle .
\lbl{4.4}
\ee
Relations among those four kinds of creation operators and wave packet profiles
are given in Eqs.\,(\ref{3.1}),(\ref{3.9}),(\ref{3.14a}),(\ref{3.14b}),(\ref{3.16}) and
(\ref{3.17}).

Let us consider the operator
\be
  \hbx = \int \dd^3 \bx \, a^\dg (\bx) \bx \, a(\bx) ,
\lbl{4.5}
\ee
which in momentum space reads
\be
  \hbx = \int \dd^3 \bp \, a^\dg (\bp) i \frac{\p}{\p \bp}\, a(\bx) .
\lbl{4.6}
\ee

The action of $\bx$ on a basis state $|\bx \rangle$ gives
\be
  \hbx |\bx \rangle = \bx |\bx \rangle .
\lbl{4.7}
\ee
The basis states $|\bx \rangle$ are thus eigenstates of the operator $\hbx$,
which can therefore be called {\it position operator}.

If we act with the operator $\hbx$ on a generic single particle state (\ref{3.18})
and make the projection onto $\langle \bx | = \langle 0| a (\bx)$ or
$\langle \bp | = \langle 0| a (\bp)$, we obtain
\be
  \langle \bx |\hbx |\Psi \rangle = \bx f(t,\bx) ,
\lbl{4.8}
\ee
\be
  \langle \bp |\hbx |\Psi \rangle = i \frac{\p}{\p \bp} f(t,\bx) ,
\lbl{4.9}
\ee
But if we project the same state (\ref{3.18}) onto the states
$\langle {\tl \bx} | = \langle 0| \tla (\bx) \equiv \langle 0| \vphi^{(-)} (0,\bx)$
or $\langle {\tl \bp} | = \langle 0| \tla (\bp)$, then we find
\be
  \langle {\tl \bx}|\hbx |\Psi \rangle 
  = \left ( \bx + \frac{\nabla}{2 (m^2-\nabla^2)^2} \right ) \tlf(t,\bx) ,
\lbl{4.10}
\ee
\be
  \langle {\tl \bp}|\hbx |\Psi \rangle 
  =  i \left ( \frac{\p}{\p \bp} - \frac{\bp}{2 \om_\bp^2} \right ) \tlg(t,\bx) ,
\lbl{4.11}
\ee
which are the well known expressions for the action of the Newton-Wigner
position operator\,\ci{NewtonWigner}--\ci{Herrmann} on a wave packet profile that satisfies
the scalar product given in Eq.\,(\ref{3.17}).

The extra term in  Eq.\,(\ref{4.11}) comes from the factor $\sqrt{(2 \pi)^3\, 2 \om_\bp}$ in
the transformation (\ref{3.1}) between $a (\bp)$ and $\tla (\bp)$, or $g(t,\bp)$ and
$\tlg (t,\bp)$. Equation (\ref{4.10}) can then be obtained from the relation (\ref{2.25})
between $\tlf (t,\bx)$ and $\tlg (t,\bp)$.

Rewritten in terms of $\tla (\bp) = \sqrt{(2 \pi)^3 2 \om_\bp}$, the position operator
(\ref{4.6}) becomes
\be
  \hbx = \int \frac{\dd^3 \bp}{(2 \pi)^3 \,2 \om_\bp} \,\tla^\dg (\bp)\,
   i \left ( \frac{\p}{\p \bp} - \frac{\bp}{2 \om_\bp^2} \right ) \tla (\bp) ,
\lbl{4.12}
\ee
where $\p/\p \bp \equiv \nabla_\bp$. Its Fourier transform
is then
\be
  \hbx = \int \dd^3 \bx \,\tla^\dg (\bx) \left ( \bx + \frac{\nabla}{2 (m^2 - \nabla^2)^2}
  \right ) \tla (\bx) . 
\lbl{4.12a}
\ee

We see that the position operator $\hbx$ has a rather cumbersome form if written
in terms of $\tla^\dg (\bp)$, $\tla (\bp)$, or $\tla^\dg (\bx)$, $\tla (\bx)$, whilst it has the
simple form (\ref{4.5}) or (\ref{4.6}) if written in terms of $a^\dg (\bp)$, $a (\bp)$,
or $a^\dg (\bx)$, $a (\bx)$. Its action on the wave packet profile in the coordinate
and the momentum representation, has the simple forms (\ref{4.8}) and (\ref{4.9}),
respectively.

The position operator in the form (\ref{4.5}) or (\ref{4.6}) is self adjoint with respect to the
scalar product (\ref{3.17}) expressed in terms of $f(t,\bx)$ or $g(t,\bp)$. The same
position operator in the form (\ref{4.12}) or (\ref{4.12a}) is self-adjoint with respect to the
scalar product (\ref{3.17}) expressed in terms of $\tlf(t,\bx)$ or $\tlg (t,\bp)$.

The representation with $a (\bx)$, $a(\bp)$, and its Hermitian conjugates, thus
gives simple expressions and it enables the interpretation of $|f(t,\bx)|^2 \equiv
f^* (t,\bx) f(t,\bx)$ as the probability density of finding a particle at position $\bx$
and time $t$.

Similarly to the position operator, we can define the momentum operator
according to
\be
  \hbp = \int \dd^3 \bp \,a^\dg (\bp) \, \bp\, a(\bp) = \int \dd^3 \bx \,  a^\dg (\bx) (-i) \frac{\p}{\p \bx} a(\bx).
\lbl{4.13}
\ee

From the commutation relations (\ref{3.3}),(\ref{3.4}), we find
\be
  [\hbx,\hbp] = i {\hat N} {\bf 1},~~~~[\hbx,\hbx']=0,~~~~[\hbp,\hbp']=0,
\lbl{4.14}
\ee
where
\be
  {\hat N} = \int \dd^3 \bx\, a^\dg (\bx) a (\bx) = \int \dd^3 \bp \, a^\dg (\bp) a(\bp)
\lbl{4.15}
\ee
is the particle number operator.

Defining the center of mass position operator,
\be
  \hbx_T = {\hat N}^{-1} \hbx ,
\lbl{4.16}
\ee
we obtain
\be
  [\hbx_T,\hbp ] = i {\bf 1}.~~~~~[\hbx_T,\hbx'_T]=0,
\lbl{4.17}
\ee
where we have used $[\hbx_T,{\hat N}] =0$ and $[\hbx_T,1]=
[\hbx,{\hat N}^{-1} {\hat N}]=0$.

If the position operator $\hbx$ acts on a state with many particles at positions
$\bx_n$ we have,
\be
  \hbx a^\dg (\bx_1) a^\dg (\bx_2)... a^\dg (\bx_N) \vac =
  (\bx_1 +\bx_2 + ... \bx_N) a^\dg (\bx_1) a^\dg (\bx_2)... a^\dg (\bx_N) \vac .
\lbl{4.18}
\ee

But if the center of mass position operator $\hbx_T$ acts on the same state,
then we have
\be
 \hbx_T \left ( \prod_{n=1}^N a^\dg (\bx_n) \right ) \vac 
 = \bx_T \left ( \prod_{n=1}^N a^\dg (\bx_n) \right ) \vac ,
\lbl{4.19}
\ee
\be
  \bx_T = \frac{1}{N} \sum_{n=1}^N \bx_n ,
\lbl{4.19a}
\ee
where we have now used the abbreviated notation for the product.

 \section{Behaviour of states under Lorentz transformations}
 
 The states $|\bx \rangle$, defined according to (\ref{4.1}) are an idealization that
 cannot be exactly realized in nature. They form the basis states in terms of which a generic single
 particle state can be expanded:
 \be
   |\Psi \rangle = \int \dd^3 \bx f(\bx) a^\dg (\bx) |0 \rangle .
\lbl{6.1}
\ee
If $f(\bx) = \delta^3 (\bx - \bx_0)$, then $|\Psi \rangle = a^\dg (\bx_0) |0 \rangle$,
but in general $|\Psi \rangle$ is a superposition (\ref{6.1}), or its many particle generalization,
\be
    |\Psi \rangle = \int \dd^3 \bx f(\bx) a^\dg (\bx) |0 \rangle 
    + \int \dd^3 \bx_1 \dd^3 \bx_2  f(\bx_1,\bx_2) a^\dg (\bx_1) a^\dg (\bx_2) |0 \rangle + ... .
\lbl{6.2}
\ee
In this paper we restrict our consideration to the single particle case, though we could
as well consider the many particle case.

When considering the behaviour of $|\Psi \rangle$, $f (\bx)$ and $a^\dg (\bx) \vac = |\bx \rangle$
under Lorentz transformations we must be careful in determining which kind of
transformation we have in mind, passive of active. In the case of a {\it passive transformation},
the state $|\Psi \rangle$ remains the same, whilst the components $f(\bx)$ and the
basis states $|\bx \rangle = a^\dg (\bx) \vac$ change.

In order to see how the expression (\ref{6.1}) for a state $|\Psi \rangle$ looks in another
Lorentz frame, let us rewrite it in terms of ${\tl f} (t,\bx)$ and ${\tl a}^\dg (0,\bx)
\equiv \vphi^\dg (0,\bx)$:
$$
|\Psi \rangle = \int \dd^3 \bx \, f(t,\bx) a^\dg (\bx) \vac =
  \int \dd^3 \bx \, \sqrt{2} (m^2- \nabla^2)^{1/4} {\tl f}(t,\bx) 
  \sqrt{2} (m^2- \nabla^2)^{1/4} {\tl a}^\dg (\bx) \vac$$
$$
  = 2 \int \dd^3 \bx \, \sqrt{m^2- \nabla^2} {\tl f}(t,\bx) {\tl a}^\dg (\bx) \vac
  = 2 i \int \dd^3 \bx \, \p_0 {\tl f}(t,\bx)  {\tl a}^\dg (\bx) \vac \hs{1cm} $$
\be
    =
  2 i \int \dd \Sigma^0 \p_0 {\tl f}(t,\bx)  {\tl a}^\dg (\bx) \vac . \hs{2cm}
\lbl{6.3}
\ee
This can be written as
\be
  |\Psi \rangle = 2 i\int \dd \Sigma^\mu \p_\mu {\tl f}(t,\bx)  {\tl a}^\dg (t_0,\bx) \vac ~,~~~t_0 =0 ,
\lbl{6.4}
\ee
where $\dd \Sigma^\mu = (\dd \Sigma^0,0,0,0)$, and ${\tl a}^\dg (\bx) \equiv {\tl a}^\dg (0,\bx)
\equiv  \vphi^+ (0,\bx)$.

The quantity ${\tl f} (t,\bx) \equiv {\tl f} (x)$ transforms under Lorentz transformations as
 a scalar, ${\tl f}' (x') = {\tl f} (x) = {\tl f} (L^{-1} x')$,
 where $x' = L x$, i.e., $x'^\mu = {L^\mu}_\nu x^\nu$.
 Similarly, also the operator $\vphi^\dg (x)$ transforms as a scalar, $\vphi'^+ (x') =
 \vphi^+ (x)$. Therefore, expressed in another Lorentz frame, the state (\ref{6.4}) reads
 \be
   |\Psi \rangle = 2 i \int \dd \Sigma'^\mu \p'_\mu {\tl f}'(t',\bx')  {\tl a}'^\dg (t'_0,\bx') \vac 
   = \int \dd \Sigma'^\mu \p'_\mu {\tl f}(L^{-1} x')  {\tl a}^\dg (L^{-1}(t'_0,\bx')) \vac .
\lbl{6.5}
\ee
Here $t'_0$ is the Lorentz transform of the time $t_0 = 0$.

In the case of a boost in the $x^1$ direction, we have
\bear
  &&\dd \Sigma^1 = 0 =  \frac{\dd \Sigma'^1 + v \dd \Sigma'^0}{\sqrt{1-v^2}} ~
  \Rightarrow ~ \dd \Sigma'^1 = - v \dd \Sigma'^0\nonumber\\
  &&\dd \Sigma^0 = \frac{\dd \Sigma'^0 + v \dd \Sigma'^1}{\sqrt{1-v^2}} =
  \frac{\dd \Sigma'^0 (1-v^2)}{\sqrt{1-v^2}} = \dd \Sigma'^0 \sqrt{1-v^2},
\lbl{6.5a}
\ear
which gives
\be
  \dd {\Sigma'}^\mu = (\dd \Sigma'^0,-v \dd \Sigma'^0,0,0) =
  \left ( \frac{\dd \Sigma^0}{\sqrt{1-v^2}}, - \frac{v \dd \Sigma^0}{\sqrt{1-v^2}},0,0 \right )
\lbl{6.5aa}
\ee
Equation (\ref{6.5}) then reads
\bear
  &&|\Psi \rangle = 2 i \int \left ( \dd \Sigma'^0 \p'_0 {\tl f}' (x') + \dd \Sigma'^1 \p'_1 {\tl f} (x')
  \right ) {\tl a}'^\dg (t'_0,\bx') \vac \nonumber\\
  &&\hs{7mm}= 2 i \int \dd \Sigma'^0 \left ( \p'_0 {\tl f}' (x') - v \p'_1 {\tl f}' (x') \right )
  {\tl a}'^\dg (t'_0,\bx') \vac .
\lbl{6.5b}
\ear
In the last expression the quantity $\dd \Sigma'^\mu = (\dd \Sigma'^0,\dd \Sigma'^1,0,0)$
represents the same hypersurface  element occurring in eq.\,(\ref{6.4}), but expressed
in a new Lorentz frame $S'$.

Instead of performing in $S'$ the integration over the same 3-surface
$\dd \Sigma'^\mu = {L^\mu}_\nu \dd \Sigma^\nu$ as in the frame $S$, in which
$\dd \Sigma^\mu = (\dd \Sigma^0,0,0,0)$, we can as well perform the integration
over a different 3-surface, whose elements are $\dd {\bar \Sigma}'^\mu = (\dd {\bar \Sigma}'^0 ,0,0,0)$,
and not those given in Eq.\,(\ref{6.5aa}).
Then, instead of (\ref{6.5}), we have a different state
$$
  |{\bar \Psi} \rangle = 2 i \int \dd {\bar \Sigma}'^0 \p'_0 {\tl f} (L^{-1} x') {\tl a}^\dg (L^{-1} ({\bar t}'_0,\bx')) \vac $$
\be
   \hs{1cm} = 2 i \int \dd {\bar \Sigma}'^0 \p'_0 {\tl f}' ( x') {\tl a}'^\dg  ({\bar t}'_0,\bx') \vac ,
\lbl{6.6}
\ee
where ${\bar t}'_0 =0$. Now ${\bar t}'_0$ is not a Lorentz transform of the time $t_0$ at a spatial position
$\bx$. The expression (\ref{6.6}) for $|{\bar \Psi} \rangle$ has the same form as the expression (\ref{6.3})
for $|\Psi \rangle$. The same steps as in Eq.\,(\ref{6.3}) can also be done in Eq.\,(\ref{6.6}).
Writing ${\tl a}'^\dg ({\bar t}'_0,\bx') \equiv {\tl a}'^\dg (\bx')$, we therefore have
\bear
  && |{\bar \Psi} \rangle = 2 i \int \dd {\bar \Sigma}'^0 \p'_0 {\tl f}' (x') {\tl a}'^\dg (\bx') \vac \nonumber \\
  && \hs{0.8cm} = 2 i \int \dd^3 \bx' \, \p'_0 {\tl f}' (t',\bx') {\tl a}'^\dg (\bx') \vac \nonumber \\
  && \hs{0.8cm} = \int \dd^3 \bx' \, 2 \sqrt{m^2 - \nabla'^2}  {\tl f}' (t',\bx') {\tl a}'^\dg (\bx') \vac \nonumber \\
  && \hs{0.8cm} = \int \dd^3 \bx' \, \sqrt{2} \,(m^2 - \nabla'^2)^{1/4}  {\tl f}' (t',\bx') 
  \sqrt{2} \,(m^2 - \nabla'^2)^{1/4}{\tl a}'^\dg (\bx') \vac \nonumber \\
  &&\hs{0.8cm} =\int \dd^3 \bx' f'(t',\bx') a'^\dg (\bx') \vac 
\lbl{6.7}
\ear

We see that in a new Lorentz frame $S'$ we can form a state $|{\bar \Psi} \rangle$ in the analogous
way as in the old Lorentz frame $S$, by using the transformed wave packet $f'(t',\bx')$ and the
transformed creation operators $a'^\dg (\bx')$. The latter operator creates a particle at the spacetime
event $(t'=0,\bx')$, whilst the original operator $a^\dg (\bx)$ creates a particle at $(t=0,\bx)$, which in
general is a different event than $(t'=0,\bx')$, and on different 3-surface.

Let us now investigate how the scalar product transforms under Lorentz transformations:
\be
  \langle \Psi|\Psi \rangle = 2 i \int_\Sigma \dd \Sigma^0 {\tl f}^* (t,\bx) \p_0 {\tl f} (t,\bx)
  = \int_\Sigma \dd^3 \bx f^* (t,\bx) f(t,\bx) .
\lbl{6.8}
\ee
Using the Lorentz transformation (\ref{6.5a}) for $\dd \Sigma^0$ and transforming $\p_0 {\tl f}$
according to
\be
  \p_0 {\tl f} (t,\bx) = \frac{\p'_0 - v \p'_1}{\sqrt{1-v^2}} {\tl f}' (t',\bx'),
\lbl{6.9}
\ee
we obtain
\bear
 &&  \langle \Psi|\Psi \rangle=2 i \int_\Sigma \dd \Sigma'^0 {\tl f}'^* (t',\bx')  
  (\p'_0 - v \p'_1) {\tl f}' (t',\bx') \nonumber \\
  &&\hs{1.4cm}= 2 i \int_\Sigma \left [ \dd \Sigma'^0 {\tl f}'^* (t',\bx')  \p'_0  {\tl f}' (t',\bx')
 + \dd \Sigma'^1 {\tl f}'^* (t',\bx')  \p'_1  {\tl f}' (t',\bx') \right ] , \ \ \ \ \  
\lbl{6.10}
\ear
where, according to (\ref{6.5a}), $\dd \Sigma'^1 = - v \dd \Sigma'^0$.
Expression (\ref{6.10}) is just a particular case of the covariant expression
\be
  2 i \int_\Sigma \dd \Sigma^\mu {\tl f}^* (x) \p_\mu {\tl f} (x) =
  2 i \int_\Sigma \dd \Sigma'^\mu {\tl f}'^* (x') \p_\mu {\tl f}' (x') ,
\lbl{6.10a}
\ee
if in the reference frame $S$ the hypersurface is $\dd \Sigma^\mu = (\dd \Sigma^0,0,0,0)$.

The scalar product is
expressed in the  frame $S$ according to Eq.\,(\ref{6.8}), and in the frame
$S'$ according to Eq.\,(\ref{6.10}). In the frame $S'$ not only the time like component,
but also the spatial component takes place. This is so because in the frame $S'$
the hypersurface element $\dd \Sigma'^\mu$, over which we integrate, has also
space like and not only time like components.

However, in every Lorentz frame we are free to choose a hypersurface over which
to perform the integration\footnote{
Frame dependent localization has bee considered in Refs\,\ci{Fleming1,Fleming2}.}.
Thus, instead of taking
$\dd \Sigma'^\mu = (\dd \Sigma'^0, - v \dd \Sigma'^0,0,0)$, which in $S$ has components
$\dd \Sigma^0,0,0,0)$, we can take another hypersurface, whose elements in the frame
$S'$ are $\dd {\bar \Sigma}'^\mu = (\dd {\bar \Sigma}'^0 , 0,0,0)$. The state is then different,
namely (\ref{6.6}), and the scalar product is then
is not that of Eq.\, (\ref{6.10}), but is
\be
 \langle {\bar \Psi}|{\bar \Psi} \rangle= 2 i \int _{{\bar \Sigma}} \dd {\bar \Sigma}'^0 \, {\tl f}'^* (t',\bx')
  \p'_0 {\tl f}' (t',\bx') .
\lbl{6.11}
\ee
The latter expression, valid in the frame $S'$, has the same form as the expression (\ref{6.8}),
valid in the frame $S$. Therefore we can proceed as in Secs\,2 and 3 and arrive at the
scalar product of the form (\ref{3.17}), and the relation (\ref{3.16}) between
$f(t,\bx)$ and ${\tl f}(t,\bx)$, in which $\dd^3 \bx$ is replaced by $\dd^3 \bx' = \dd {\bar \Sigma}'^0$,
${\tl f} (t,\bx)$ by ${\tl f}' (t',\bx')$ and $f(t,\bx)$ by $f'(t',\bx')$. Therefore, the scalar product (\ref{6.11})
can be written in the form
\be
   \langle {\bar \Psi}|{\bar \Psi} \rangle = \int \dd^3 \bx' f'^* (t',\bx') f' (t',\bx') ,
\lbl{6.12}
\ee
where $f'^* (t',\bx') f' (t',\bx')$ is the probability density in the new Lorentz frame.

Despite that the integrals
in Eqs.\,(\ref{6.11}) and (\ref{6.12}), or in Eq.\,(\ref{6.8}),  are equal, the expressions under the integrals, are not equal\,\ci{Horwitz}. For an illustrative discussion see Refs.\,\ci{Nikolic1,Nikolic2}.

\section{Wave packet solutions of the relativistic Schr\"odinger equation}

We have seen that a wave packet profile $f(t,\bx)$ for a single particle state, created by
the action of $a^\dg (\bx)$ on the vacuum, satisfies the relativistic Schr\"odinger equation
(\ref{3.10a}). Recall that we have obtained such equation within the framework
of relativistic quantum field theory (QFT). Usually Eq.\,(\ref{3.10a}) is considered
from the point of view of {\it relativistic quantum mechanics} (QM). 
But the straightforward procedure, displayed in this paper,  (see also Refs.\,\ci{Horwitz,Karpov1})
has shown that relativistic QM (restricted to positive norms) is embedded
within relativistic QFT, namely,  it is associated with single particle wave packet
profiles that, as shown in Sec.\,2, automatically have positive norms and energies,
once a vacuum, satisfying $\tla (\bk) \vac =0$, is chosen. 

We will now study wave packet solutions of equation (\ref{3.10a}). Let initially the
wave function be given by
\be
  f(0,\bx) = \delta^3 (\bx - \bx_0),
\lbl{5.1}
\ee
its Fourier transform being
\be
  g(0,\bp) = \frac{1}{\sqrt{(2 \pi)^3}} \, {\rm e}^{i \bp \bx_0} .
\lbl{5.2}
\ee
The latter state evolves according to Eq.\,(\ref{3.7}), which gives
\be
  g(t,\bp) = {\rm e}^{-i \om_\bp t} g(0,\bp).
\lbl{5.3}
\ee
A single particle state is thus
\be
  |\Psi (t) \rangle = \int \frac{\dd^3 \bp}{\sqrt{(2 \pi)^3}}\, {\rm e}^{-i \om_\bp t + i \bp \bx_0}
  a^\dg (\bp) \vac ,
\lbl{5.4}
\ee
its projection onto a state $\langle \bx | = \langle 0| a (\bx)$ being the Green's
function
\bear
  &&\langle \bx |\Psi (t) \rangle 
  = f(t,x) = \frac{1}{\sqrt{(2 \pi)^3}} \int \dd^3 \bp \, {\rm e}^{-i \sqrt{m^2 + \bp^2}t}
  {\rm e}^{i \bp (\bx - \bx_0)} \nonumber \\
  &&\hs{3.3cm} = \frac{1}{\sqrt{(2 \pi)^3}} {\rm e}^{-i \sqrt{m^2 -\nabla^2} t} \delta^3 (\bx-\bx_0)\nonumber \\
  &&\hs{3.3cm}  =G(t,\bx;0,\bx_0) .
\lbl{5.5}
\ear

For a generic initial wave function $f(0,\bx)$, different
from (\ref{5.1}), we have
\be
  f(t,\bx) = \int \dd^3 \bx' \, G(t,\bx;0,\bx') f(0,\bx').
\lbl{5.6}
\ee
As explicitly derived in Ref.\,\ci{Al-Hashimi}
(see also\,\ci{Cirilo-Lombardo}), the Green function {\it in one dimension} is
\be
  G(t,x;0,0) \equiv G(t,x) 
   = - \frac{i m t}{\pi \sqrt{x^2 - t^2}}\, 
   K_1 \left ( m \sqrt{x^2 - t^2} \right ) ,
\lbl{5.7}
\ee
where $K_1$ is the modified Bessel function of degree one.
Equation (\ref{5.7}) is valid for all values of $x$ and $t$.
\setlength{\unitlength}{.8mm}

\begin{figure}[h!]
\hs{3mm} \begin{picture}(120,70)(0,0)
\put(5,0){\includegraphics[scale=0.4]{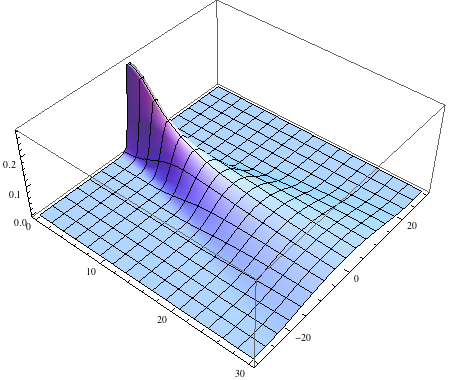}}
\put(100,0){\includegraphics[scale=0.48]{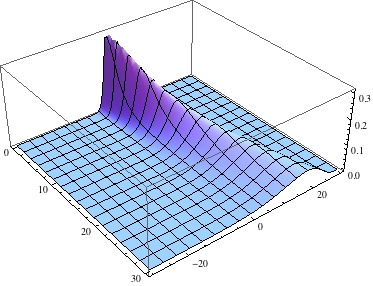}}

\put(7,59){$\langle v \rangle=0$}
\put(100,59){$\langle v \rangle =0.5$}
\put(80,7){$\Delta x = 1.4$}

\put(-2,37){$|f|^2$}
\put(92,37){$|f|^2$}

\put(22,14){$t$}
\put(112,14){$t$}

\put(69,16){$x$}
\put(156,10){$x$}

\end{picture}

\caption{\footnotesize Evolution of the probability density, $|f|^2$, for a minimal position-velocity
uncertainty wave packet, $f(t,x)$, whose width is $\Delta x > \lambda_c = 1/m$, for two different
velocities $\langle v \rangle$.  We can express $m$ in arbitrary units, therefore we take $m=1$.}
\end{figure}
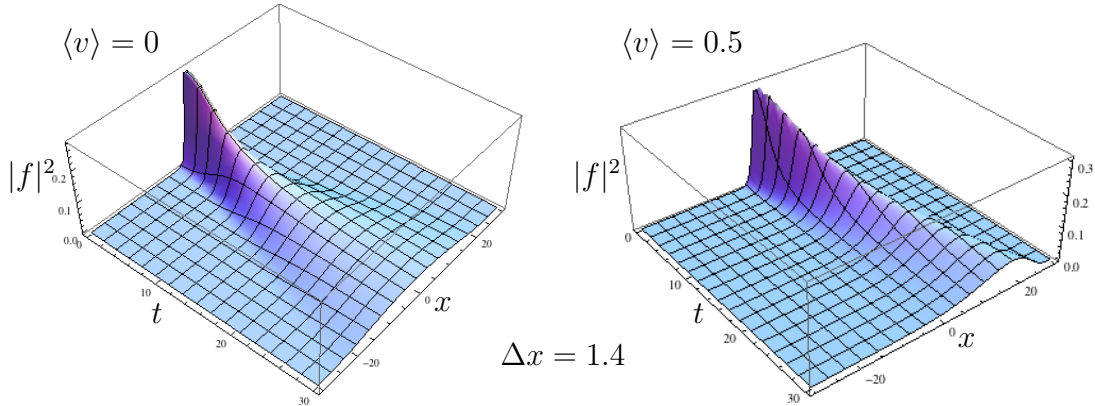 

Al-Hashimi and Wiese\,\ci{Al-Hashimi} also showed that the
wave function of a minimal position-velocity uncertainty
wave packet can be expressed in terms of the Green function
according to
\be
  f(t,x) = A G (x- i \beta,t-i \alpha) ,
\lbl{5.8}
\ee
where $A$ is a normalization constant, and where
$\alpha$, $\beta = \beta_r + i \beta_i$ are constants,
related to the wave packet parameters according to
\be
  \alpha = \frac{1}{2 \Delta v^2} \langle \p_p^2 E \rangle~,
    ~~~~\beta_r = \alpha \langle v \rangle~,~~~~
    \beta_i = - \langle x \rangle ,
\lbl{5.9}
\ee
where $\Delta v^2 = \langle v \rangle^2 - \langle v^2 \rangle$.
Taking into account the relation for a minimal
position velocity uncertainty wave packet\,\ci{Al-Hashimi},
\be
  \Delta x \Delta v = \frac{m^2}{2 \langle E \rangle^3} ,
\lbl{5.10}
\ee
we find
\be
  \alpha = \frac{2 \Delta x^2 m}{\left ( 1 - \langle v \rangle^2  \right )^{3/2} }.
\lbl{5.11}
\ee
\setlength{\unitlength}{.8mm}

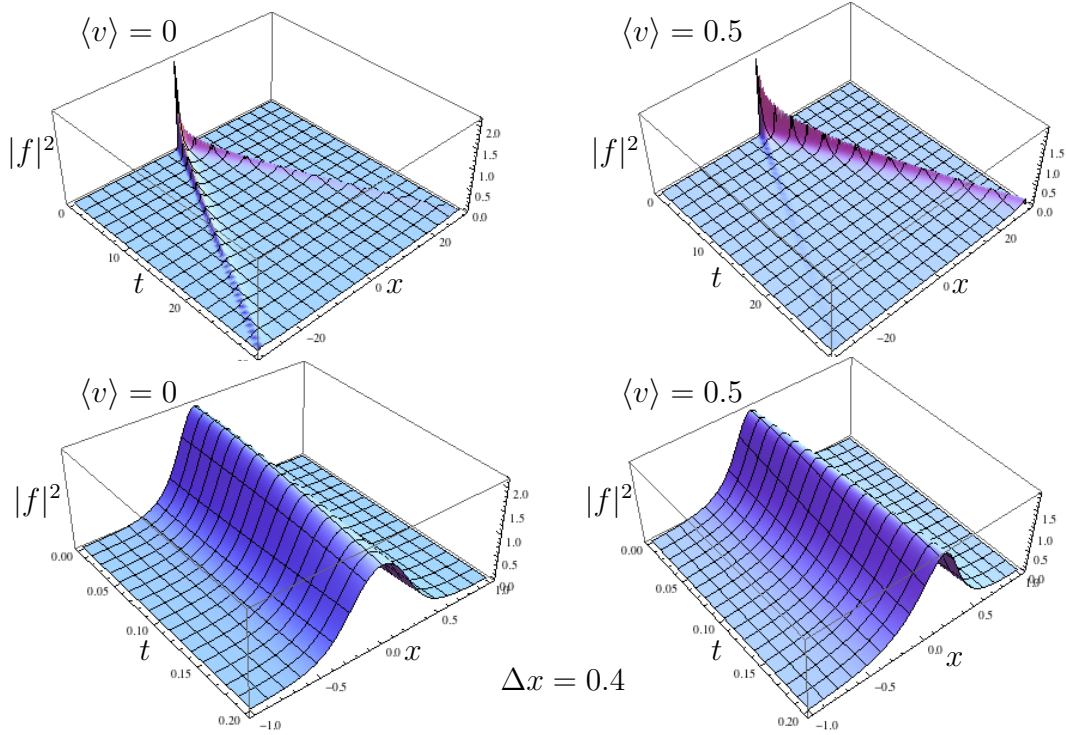
\begin{figure}[h!]
\hs{3mm} \begin{picture}(120,120)(0,0)
\put(5,60){\includegraphics[scale=0.45]{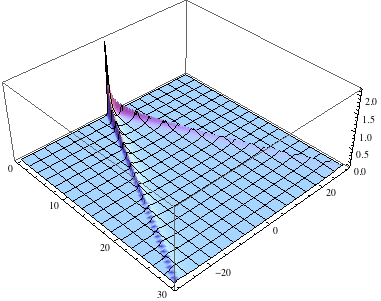}}
\put(100,60){\includegraphics[scale=0.45]{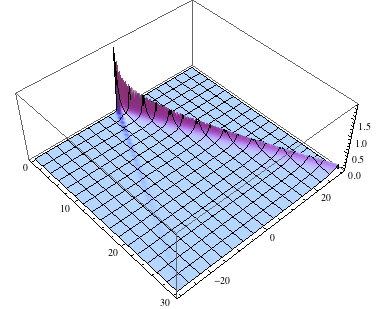}}
\put(5,0){\includegraphics[scale=0.45]{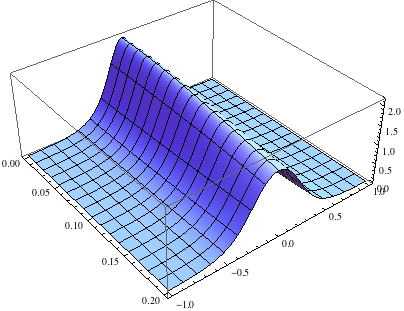}}
\put(100,0){\includegraphics[scale=0.45]{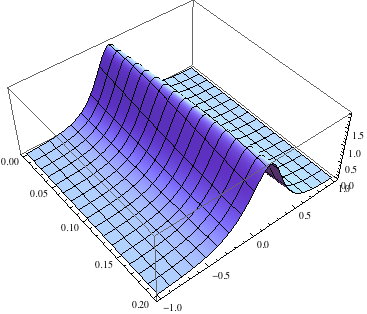}}

\put(10,115){$\langle v \rangle=0$}
\put(100,115){$\langle v\rangle =0.5$}
\put(10,55){$\langle v \rangle =0$}
\put(100,55){$\langle v \rangle =0.5$}
\put(80,7){$\Delta x = 0.4$}

\put(-1,37){$|f|^2$}
\put(94,37){$|f|^2$}

\put(20,12){$t$}
\put(115,12){$t$}

\put(64,12){$x$}
\put(154,11){$x$}
\put(-2,95){$|f|^2$}
\put(95,95){$|f|^2$}

\put(19,73){$t$}
\put(115,74){$t$}

\put(61,73){$x$}
\put(155,73){$x$}

\end{picture}

\caption{\footnotesize Evolution of the probability density, $|f|^2$, for
a minimal position-velocity uncertainty wave packet, $f(t,x)$,
whose width is $\Delta x < 1/m$, for two different values of $\langle v \rangle $.  Initially the wave packet evolves
normally (lower plots), but after certain time it splits into two branches (upper plots).}

\end{figure} 

Using Mathematica we have calculated the probability density
$|f(t,x)|^2$ for various choices of parameters $m$, $\Delta x$
and $\langle v \rangle$. For the parameter $\beta_i$, which
determines the initial position of the wave packet, we have
set $\beta_i =0$. In Fig.\,1 are shown the plots for $\langle v \rangle =0$, $\langle v \rangle =0,5$,
$m=1$, and the wave packet width $\Delta x=1.4$.
We see\footnote{
We use the extended Planck units\,\ci{PavsicBook} (see also Wikipedia\,\ci{WikiPlanck}) in which $\hbar = c=G= 4 \pi \epsilon_0 = 1$.}
that for $\Delta x > \lambda_c$,
where $\lambda_c = \frac{\hbar}{m c}= \frac{1}{m}$
is the Compton wavelength, we have just the usual wave packet solution
with the maximum of $|f(t,x)|^2$
 corresponding to the expectation value of the
particle's classical trajectory (Fig.\,1). But if 
$\Delta x < \frac{1}{m} =1$, then during certain period the
wave packet evolves normally, and afterwhile it splits into
two wave packets, whose centers move into the opposite directions
with the velocity of light. The information about the wave packet
expectation velocity is encoded in different intensities of the two
branches (Fig.\,2).
\setlength{\unitlength}{.8mm}

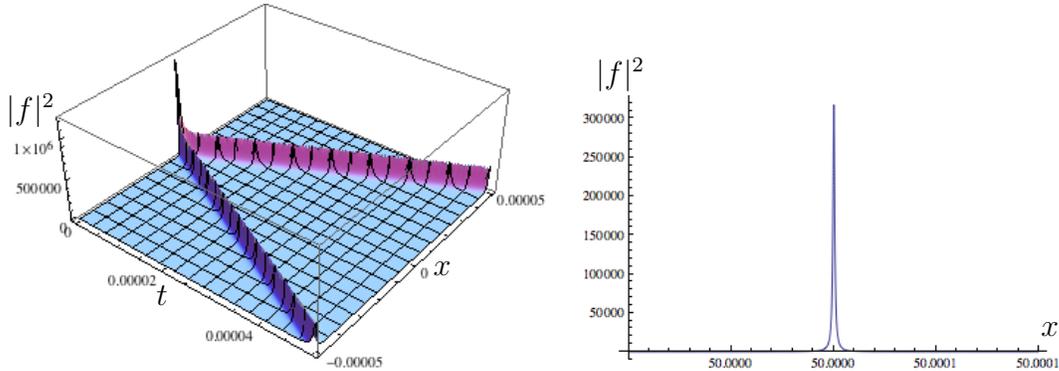
\begin{figure}[h!]
\hs{3mm} \begin{picture}(120,60)(0,0)
\put(5,0){\includegraphics[scale=0.55]{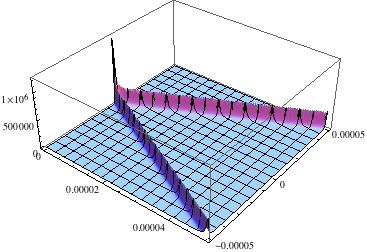}}
\put(100,0){\includegraphics[scale=0.5]{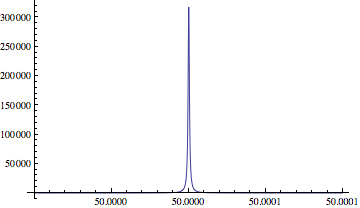}}

\put(4, 42){$|f|^2$}
\put(102,48){$|f|^2$}

\put(29,11){$t$}
\put(75,16){$x$}
\put(176,6){$x$}

\end{picture}

\caption{\footnotesize 
An example of the probability density, $|f|^2$, for a wave packet, whose initial width, $\Delta x = 0.0005$,
is very small in comparison with the Compton length, which in our units is  $\lambda_c =1$.
We see that the probability density with decreasing $\Delta x$ becomes more and more concentrated on
the light cone.}
\end{figure} 

Inspecting the wave packets of Figs.\,1 and 2, it is obvious that
when observed from another Lorentz frame nothing unusual happens.
In another Lorentz frame they become Lorentz transformed wave packets.
If the initial width decreases, then the probability density $|f|^2$ becomes higher
and higher, as shown in Fig.\,3. In the limit of a $\delta$-like localized wave packet
at $t=0$, $|f|^2$ becomes infinitely high and infinitely narrow, concentrated on
the light cone, according to $|f|^2 = \frac{1}{2} \left (\delta(t-x)+\delta (t+x) \right )$. The event at $t=0$ and $x=0$ at which the particle is initially localized,
is, of course, invariant in all Lorentz frames. Thus all observers see the particle localized
in the origin of their Lorentz frame. At later times $t>0$ the particle is localized on the
intersection of the simultaneity hypersurface with the light cone. For such a limiting
state, their is no instantaneous spreading of the probability density of the sort
considered in Refs.\,\ci{Mosley,Rosenstein,Hegerfeldt}. We thus see that the relativistic wave packet
in the limit of the $\delta$-like initial localization in fact remedies the non relativistic
case, in which an infinitely thin wave packet, exactly localized at $t=0$, spreads
over all space at arbitrarily small $t>0$.

We have also seen that the relativistic expression (\ref{5.8}), derived from (\ref{5.5}),
describes wave packets of any velocity, including zero velocity. Thus even a particle
moving with zero velocity is described by the {\it relativistic wave packet}. The non
relativistic wave packet is obtained from expression (\ref{5.6}) in the
approximation $m^2 \gg \bp^2$ in which we neglect higher momenta. Equivalently, it is obtained from
expression (\ref{5.8}) if the wave packet width $\Delta x$ is large in comparison with the Compton length.

The case in which at $t=0$ a wave is not a minimal
position-velocity wave packet, but an exactly localized (rectangular) wave packet,
was considered by Karpov et al.\,\ci{Karpov1}.  It was found that such wave packet
is a superposition of two non local wave packets moving in the opposite directions
with the velocity of light. Initially this gives a rectangular localized wave packet, which immediately
delocalizes at $t>0$. This is similar to the behavior of a minimal position-velocity
wave packet, whose width $\Delta x$ is smaller than the Compton wavelength, with the
difference that the separation into two distinct wave packets becomes manifest
imediately, and not after certain period. Such exact initial localization (as a rectangular wave packet), of course, is
not invariant under Lorentz transformations. When observed from another frame, the simultaneity
hypersurface $\Sigma'$ is no longer the same, and it intersection with the evolving wave packet
does not give an exact localization on $\Sigma'$, but a localization with an infinite tail. The exception,
as we have seen above, is
the limiting case when the width of the exact localization goes to zero and we approach the localization
at a spatial point. Such, initially $\delta$-like localized, wave packet does not instantly evolve into a wave packet
with infinite tail, but remains localized on the light cone.

\section{On the causality violation of a relativistic wave packet}

Inspecting the wave packet in Fig.\,2 one observes that the probability density extends
accross the light ``cone". Many authors have analysed such behaviour in view of a possible
causality violation\,\ci{Hegerfeldt,Hegerfeldt2,Rosenstein,Mosley,Wagner,Eckstein,Barat,Buscemi}.
However, causality would be violated if {\it information} could be sent
faster than light. The fact that some part of the {\it probability} density arrives at a position
$x$ earlier than light, by itself does not guarantee that information can also arrive quicker
than light. With a single particle one cannot send information, because the position $x$
at which the particle will be detected is uncertain. One needs a modulated particle beam,
e.g., a sequence of pulses of many particles, a statistical mixture of them. Then the sum
$\sum_i|f_i|^2$ is proportional to the density $\rho$ of particles at a position $x$ at a time $t$. In
Fig.\,4 it is shown how the density at the fixed position $x=1$ changes with time in the
situation in which after a first wave packet $f_1$, formed at $t=0$, a second, similar, wave
packet $f_2$, formed at $t=0.1$, is emitted.
In the right plot both densities are summed. We see that at $t<1$ there is no modulation
of the particle density, which indicates that in such an arrangement information cannot
be transmitted faster than light. The fact that $\rho$ starts to increase before $t=1$, which in this
units is the arrival time of light,
does not automatically imply that a message has been received at $t<1$, because at
that earlier time there has been no obvious modulation of the density $\rho$. 

\setlength{\unitlength}{.8mm}

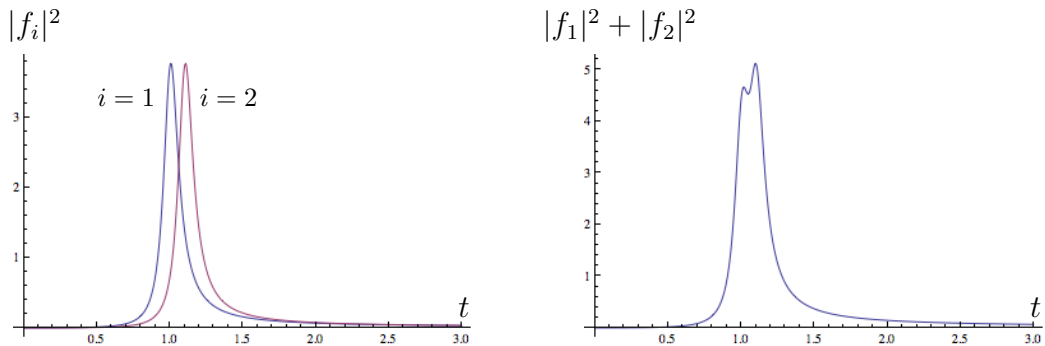
\begin{figure}[h!]
\hs{3mm} \begin{picture}(120,60)(0,0)
\put(5,0){\includegraphics[scale=0.48]{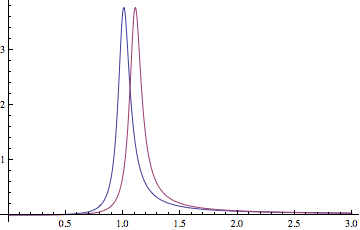}}
\put(100,0){\includegraphics[scale=0.48]{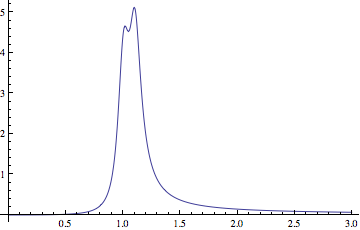}}

\put(4,52){$|f_i|^2$}
\put(93,52){$|f_1|^2+|f_2|^2$}
\put(19,40){\footnotesize $i=1$}
\put(36,40){\footnotesize $i=2$}
\put(79,5){$t$}
\put(174,5){$t$}

\end{picture}

\caption{\footnotesize Time dependence of the probability density $|f_i|^2$, $i=1,2$,
and their sum, observed at a fixed position $x=1$
for two subsequent wave packets with $\Delta x = 0.2$ and $v=0$. The first wave packet
is formed at $t=0$, and the second one at $t=0.1$.}
\end{figure} 

Alternatively, a beam of particles can be modulated spatially, e.g., by an arrangement of slits,
and so bear a message or
a signal. A possible  setup is shown in Fig.\,5 in which
the wave packet wavelength $\lambda$ is small enough, so that the packet can go through any of the slits
${\cal S}$
more or less undisturbed. An alternative arrangement is shown in Fig.\,6, where
$\lambda$ is great enough for diffraction and interference effects to occur, so that spherical wave
packets emerge from the hole and then interfere on the arrangements of slits ${\cal S}$.
If the width
of the wave packet is smaller than the Compton wavelength, then the
message comes to the detectors faster than light. Because a superluminal effect
of the wave packet is effectively observable within the Compton wavelength $\lambda_c$,
the arrangement of detectors ${\cal D}$ should be within a distance $L < \lambda_c$. In order to be able
to send a message into the past, the arrangement ${\cal D}$ should move with an
appropriate velocity (see Refs.\,\ci{Fleming2}). Moreover, such a message would
arrive into the very nearby past (within the time that takes light to travel the distance
$L<\lambda_c$).
\setlength{\unitlength}{.8mm}

 \begin{figure}[h!]
\hs{3mm} \begin{picture}(120,65)(0,0)
\put(40,0){\includegraphics[scale=0.48]{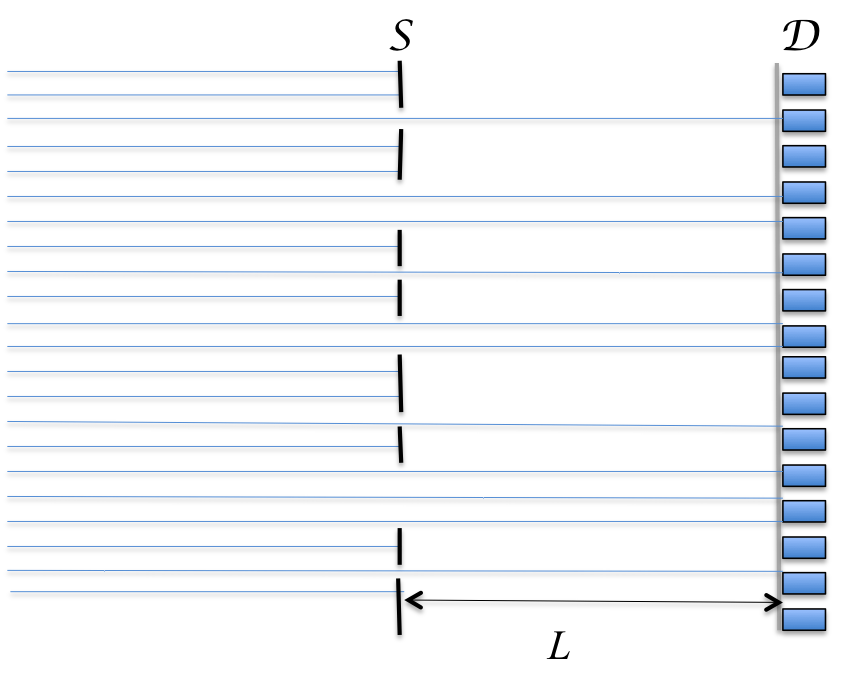}}

\end{picture}

\caption{\footnotesize Experimental setup for transmission of a signal by means of a
spatially modulated beam of particles on the arrangements of slits ${\cal S}$, each particle
being described as  a wave packet.}
\end{figure} 

We see that by using relativistic wave packets, we apparently cannot violate causality
on the macroscopic level, because the experimental setups of Figs.\,4 and 5 have
either difficult to achieve or contradictory constraints.
A mere look at Fig.\,2, in which the width of the wave packet
(and hence its superluminal tail) is smaller than $\lambda_c \sim 10^{-15} m$, reveals
that causality, in the sense of sending a signal into a reasonably remote past,
cannot be so easily violated, if at all. A very ingenious experimental setup would be
necessary for a macroscopic observer being able to invoke causality violating
situations \`a la ``grand father paradox'' or its simpler versions in which the
apparatus is destroyed before emitting a signal. Even then, causality would be
restored within a proper quantum mechanical description of the situation\,
\ci{PavsicLettNuovCim,Deutsch1991,Deutsch1,PavsicBook}.
\setlength{\unitlength}{.8mm}

\begin{figure}[h!]
\hs{3mm} \begin{picture}(120,60)(0,0)
\put(40,0){\includegraphics[scale=0.48]{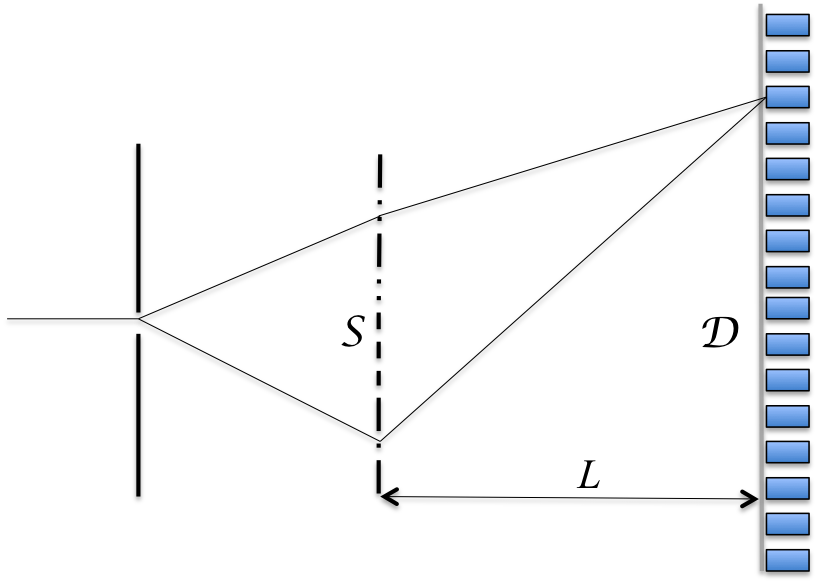}}

\end{picture}
\caption{\footnotesize An alternative experimental setup for transmission of a signal by means of a
spatially modulated beam of particles. In this setup, a beam of wave packets whose wave length
$\lambda$ and the spread $\Delta \lambda$ is greater than the diameter of the hole so that
spherical wave packets emerge from the hole and interfere on the modulated arrangements
of slits ${\cal S}$.}
\end{figure} 

The above reasoning indicates that the issue of causality violation of relativistic wave
packets is not so straightforward as it is usually assumed. Also Fleming\,\ci{Fleming2}
and Wagner\,\ci{Wagner} have
come to a similar conclusion. Ruijsenaars\,\ci{Ruijsenaars} has pointed out that the detection of `acausal events' is
vanishingly small under present laboratory conditions.
Eckstein and Miller\,\ci{Eckstein}
observed that ``causality brekdown'' has a transient character which, according
to our finding, does not automatically imply the possibility information transmission into
the past. Karpov et al.\,\ci{Karpov1} pointed out that the two complex components forming
an initially localized wave packet move causally, with the velocity of light in the opposite directions.
In their example the initial wave packet of a massless particle was localized within a rectangle, and
afterwards it had the long tails that decayed with the distance $x$ according to $b/x$  for
$b/x << 1$, where $b$ was the size of the localized wave packet. They wrote\,\ci{Karpov1}:
\begin{quotation}
{\small [Such long tails] are precursors to the usual wave propagation. Although we may have instant
interactions, these are not result of superluminal propagation, but of ``preformed'' structures.}
\end{quotation}
Further, Antoniou et al.\,\ci{Karpov2} considered a quantum electrodynamics case and demonstrated
the appearance of nonlocal effects at the level of states. They showed that the expectation value
of the electromagnetic field spreads causally, and that the classical measurements cannot detect 
the ``acausal'' effects of this non-locality. 

In this connection let me point out that with waveguides one can arrange
setups in which the group velocity of waves, the so called evanescent waves,
 is greater than the velocity of light (see e.g., \,\ci{Nimitz,Recami}). There has been a lot of
 discussion about whether or not such evanescent waves can transmit information
 faster that light. Many authors agree that in such cases the group velocity is
 not the velocity of information transmission, and that information travels slower
 than light. But Nimitz\,\ci{Nimitz} has shown that signals in such arrangements are
 indeed superluminal, and yet they do not violate causality in the sense
 that the effect cannot precede the cause. This is so, because a signal has a finite
 duration. Therefore, in a typical setup in which an observer $A$ sends a superluminal pulse-like signal to 
 a fast moving observer $B$, the pulse-like signal sent back from $B$ to $A$, because of the pulse's finite
 width, cannot arrive into the past of $A$.
  
 In the case of evanescent waves a faster than light group velocity
 does not automatically imply causality violation. We have seen that also the existence
 of superluminal tails in relativistic wave packets does not automatically imply the
 possibility of superluminal communication and thus causality violation. Moreover, in previous
 section we have demonstrated that if an effective width of  a wave packet goes to zero, then the
 probability density approaches the exact localization on the light cone. A wave packet behaves
 apparently ``acausally'' only if its width $\Delta x$ is smaller than the Compton wavelength $\lambda_c$, 
 but if $\Delta x$ goes to zero, then the ``acausal'' behavior disappears.
  
 A deeper and more
detailed thorough analysis has to be done, before we can say for sure  that causality in the
sense of ``the grand father paradox''
can be violated with relativistic wave packets. And if it is apparently violated, then we should seek how to
remedy the situation, and not reject prematurely the concept of relativistic wave packets. We have seen
that relativistic wave packets, either in momentum or in position space,
are unavoidable ingredients of relativistic quantum field theory, and may represent various types of a particle's localization, including point-like, rectangular, or Gaussian-like localization.

\section{Conclusion}

We have clarified the well known difficulties regarding localization of states in
relativistic quantum mechanics and quantum field theories. For this purpose we
proceeded step by step and thus more or less reviewed certain known
facts and results of quantum field theory, which enabled us to avoid some loopholes
and point to  connections that have been usually overlooked in the treatments that
considered only a part of the full story.

In quantum field theory  the basis states of the Fock space are created by
the action of field operators on a vacuum. In order to obtain a generic state, one has to superpose
such basis states by means of a wave packet profile (wave function) which, in general, is complex valued.
It satisfies the Schroedinger equation
with the Hamilton operator, which is positive definite with respect to the so defined Fock space states.
Only positive frequencies occur in the
{\it wave function}, whilst the {\it field operators}, expanded in terms of creation and
annihilation operators,  contain both positive and negative frequencies. A complex valued wave function
should not be confused with a non hermitian field (operator). Therefore, even in the case of a hermitian
field (operator), the corresponding wave function can be complex.

On a wave function and basis states one can apply a suitable functional transformation\,\ci{Foldy,Fonda,Horwitz},
such that the state remains the same. So we can transform the Klein-Gordon wave function
into a new wave function, called\,\ci{Horwitz} the Newton-Wigner-Foldy wave function, 
whose absolute square gives the probability density, either in momentum
or position space. Similarly, we can transform the Klein-Gordon creation operators
into new operators that create eigenstates of the Newton-Wigner position operators,
i.e., Newton-Wigner localized states. 

The type of localization is determined by the shape of of the wave function. It can be (i) point-like
localization, or  (ii) localization in a finite region of space vanishing outside, or (iii) localization in
a finite region decaying with power or exponential law, or (iv) ``effective'' localization like a Gaussian
wave packet. Usually, by ``localization"  is understood the localization of the type (i) or (ii), but in this paper we
use  the word ``localization'' for the localization of the type (iii) or (iv) as well.
The wave packets and the corresponding probability currents can be transformed from one to
another Lorentz frame. We have found that nothing unusual  happens with the wave packets
when observed from different Lorentz frames. A state localized around a certain position, remains
localized in another frame as well. This is consistent with the fact that if we observe
a wave packet in its rest frame, then it behaves approximately as a non relativistic wave packet which can be
localized. If we observe the same wave packet from a moving frame, it remains localized. The very
existence of particle pulses in accelerators confirms that even fast moving particles can be
localized. However, a state, initially localized according to (ii) in a frame $S$, is localized
according to (iii) if observed from another Lorentz frame $S'$. This is so, because simultaneity
is not invariant and because the type (ii)
localization at $t=t_0$ in the frame $S$ is only momentary, immediately switching at any later time $t>t_0$ 
to the type (iii) localization. A special case is type (i) localization at $t=t_0$, which is a limiting case
of the type (iv) localization when the width $\Delta x$ of a Gaussian-like wave packet approaches to zero.
We have demonstrated that
at later times the probability density is given by $|f(t,x)|^2$ which approaches to
$|f|^2 = \frac{1}{2} \left (\delta(t-x)+\delta (t+x) \right )$ if
$\Delta x$ goes to zero. In the limiting case of a point-like initial localization the particle is thus
localized on the light cone at any later instant. Also when observed from another frame, the particle
remains localized on the light cone. The initial point-like localization is Lorentz invariant.

 Despite that the wave packets whose width is smaller than the Compton
wave length leak outside the light cone, they cannot be used for faster than light communication
between {\it macroscopic} observers and devices.
A transfer of information cannot be done with a single wave packet, but requires, e.g., modulated in time sequences of wave packets, which move at most with the velocity of light. On the contrary,
a spatially modulated bunch of particles, localized within their Compton wavelength, can bring a
signal or a message with a superluminal velocity to a position within the Compton length from the
source. But the Compton wavelength of elementary
particles is around $10^{-15}$\,m (for electron) or smaller, so that an observer, even if by
an ingenious way could send information (message, signal) into the past, that past would be
only about $10^{-23}$\,s from his present, so that no causality paradox of the ``grand father
paradox'' or similar, could take place. But even if by an ingenious technology, creation of 
apparently paradoxical  situations were possible at the macroscopic level, there would remain
a possibility to explain them\,\ci{PavsicLettNuovCim,Deutsch1991,Deutsch1,PavsicBook}
within an appropriate quantum setup\,\ci{Everett1,Everett2,Everett3,DeWitt1,DeWitt2,Deutsch2,Zeh}.

We conclude that the usual arguments against localized relativistic states can be circumvented.
Such states naturally occur within quantum field theory and are not problematic at all.
This sheds new light on the implications of the Reeh-Schlieder theorem\,\ci{Reeh-Schlieder},
which is interpreted as implying that states (including single particle states) cannot be exactly localized
in a finite region (see, e.g.,\,\ci{Al-Hashimi}). Such a conclusion comes from the fact that one
of the axioms of algebraic quantum field theory\,\ci{Haag} is causality. However,
as pointed out by G. Valente\,\ci{Valente}, one has to distinguish among different concepts
of `causality' used in the literature, and not all imply the possibility of information transmission.
Moreover, Karpov et al.\,\ci{Karpov1} and Antoniou et al.\,\ci{Karpov2}
have demonstrated that the classical measurement cannot detect the ``acausal" effects of the wave packet
quantum states. 
In the scenario that occurred in the Reeh-Schlieder theorem, the superluminal influence of a field in
one spacetime region to a field in another region cannot be used for a controlled transmission
of information.  Therefore, the Reeh-Schlieder theorem does not imply that quantum states cannot
be localized in a finite region. They can be localized, but their immediate spreading
over all the space, cannot be used for a superluminal transmission of information.

\vs{6mm}


{\bf \large Appendix A: The propagator}

\vs{2mm}

The scalar product of two states (\ref{3.18}) at different times can be expressed as
\bear
  &&\langle \Psi_2 (t')|\psi_1 (t) \rangle = \langle 0| \int a(\bx') f_2^* (t',\bx') \dd^3 \bx' 
  \dd^3 \bx f_1 (t,\bx) a^\dg (\bx) \vac \nonumber\\
  &&\hs{2.5cm} =  \langle 0| \int \tla(\bx') 2 \om_{\bx'} \tlf_2^* (t',\bx') \dd^3 \bx' 
  \dd^3 \bx \, 2  \om_{\bx}\tlf_1 (t,\bx) \tla^\dg (\bx) \vac ,
\lbl{A1}
\ear
where $\om_{\bx} \equiv \sqrt{m^2 - \nabla^2}$. Using the Schr\"odinger equation (\ref{2.30})
and (\ref{3.10a}), we have
\be
  \tlf (t,\bx) = {\rm e}^{-i \om_\bx t} \tlf (0,\bx) ,\lbl{A1a} \ee
\be  f (t,\bx) = {\rm e}^{-i \om_\bx t} f (0,\bx) ,\lbl{A1b} .
\ee

The initial and final wave packet profiles $f_{1,2} (0,\bx)$ or $\tlf_{1,2} (0,\bx)$ are arbitrary.
Let us consider two choices:
\be
   (i) ~~~~~~~f_1(0,\bx) = \delta (\bx - \bx_0)~,~~~~f_2(0,\bx') = \delta (\bx' - \bx'_0) \hs{1cm}.
\lbl{A2}
\ee
Then we obtain
  $$\langle \Psi_2 (t')|\psi_1 (t) \rangle = \langle 0|  {\rm e}^{i \om_{\bx'_0} t'} a(\bx'_0) 
  {\rm e}^{-i \om_{\bx_0} t} a^\dg (\bx_0) \vac$$
\be
  \hs{1cm} = {\rm e}^{i \om_{\bx_0} (t'-t)} \delta (\bx'_0 - \bx_0) ,
\lbl{A3}
\ee
which is just the Green function (\ref{5.5}). 
\be
   (ii) ~~~~2 \om_\bx \tlf_1(0,\bx) = \delta (\bx - \bx_0)~,~~~~2 \om_\bx' \tlf_2(0,\bx') = \delta (\bx' - \bx'_0) ,
\lbl{A4}
\ee
then
 $$\langle \Psi_2 (t')|\psi_1 (t) \rangle = \langle 0|  {\rm e}^{i \om_{\bx'_0} t'} \tla(\bx'_0) 
  {\rm e}^{-i \om_{\bx_0} t} \tla^\dg (\bx_0) \vac$$
\be
  \hs{1.8cm} = {\rm e}^{i \om_{\bx_0} (t'-t)}\frac{1}{2 \om_{\bx_0}} \delta (\bx'_0 - \bx_0) ,
\lbl{A5}
\ee
where we have used $[\tla (\bx'),\tla^\dg (\bx)] = (1/(2 \om_\bx)) \delta (\bx' - \bx)$.
Because $\tla^\dg (\bx) \equiv \vphi^+ (0,\bx)$,  $\vphi^+ (t,\bx) = {\rm e}^{-i \om_\bx t} \vphi(0,\bx)$,
we can write Eq.\,(\ref{A5}) in the form ($x \equiv (t,\bx)$)
\be
  \langle \Psi_2 (t')|\psi_1 (t) \rangle = \langle 0|\vphi(x') \vphi^+ (x) \vac ~, ~~~~t' > t .
\lbl{A6}
\ee
If we do not impose the condition $t' > t$, then the right hand side of Eq.\,(\ref{A6}) can be written
in terms of the time ordered product $\langle 0|T\vphi (x') \vphi (x) \vac$, which is the usual QFT
propagator.

Both propagators, (\ref{A3}) and (\ref{A5}) (i.e., (\ref{A6})), are special cases of the scalar product (\ref{A1})

In the case (i), the initial and final wave packet profiles are localized according to (\ref{A2}).
This is the localisation studied in this paper. The initial, and analogously the final, state are then of the form
\be
  |\psi (0) \rangle = \int \dd^3 \bx f(0,\bx) a^\dg (\bx) \vac = a^\dg (\bx_0) \vac \equiv |\bx_0 \rangle,
\lbl{A7a}
\ee
and the scalar product (\ref{A1}) gives (\ref{A3}), which can be written as
\be
  G(t',\bx';t,\bx) = \langle \bx' |{\rm e}^{i H(t'-t)} |\bx \rangle ,
\lbl{A7b}
\ee
where the Hamilton operator in the $\bx$ representation is $\om_\bx = \sqrt{m^2 - \nabla^2}$.
Using (\ref{3.16a}), the same localized state $f(0,\bx)= \delta^3 (\bx - \bx_0)$ can be expressed in terms of
the functions $\tlf (0,\bx)$ as
\be
  \tlf (0,\bx) = \sqrt{2 \om_\bx} f (t,\bx) \equiv \sqrt{2 (m^2 - \nabla^2)}\, \delta^3 (\bx - \bx_0),
\lbl{A7}
\ee

In the case (ii), the initial  wave packet (and analogously the final wave packet) is determined by (\ref{A4}),
so that
\be
  |\psi (0) = \int \dd^3 \bx \, 2 \om_\bx \tlf (0,\bx) \tla^\dg (\bx)\vac = \tla^\dg (\bx_0) \vac \equiv |{\tl \bx_0)} \rangle.
\lbl{A8a}
\ee
The scalar product (\ref{A5}) can then be written in the form
\be
  {\tl G} (t',\bx';t,\bx) = \langle {\tl \bx}'|{\rm e}^{i H (t' -t)} |{\tl \bx} \rangle.
\lbl{A8b}
\ee

We have thus two kinds of propagators, (\ref{A7b}) and (\ref{A8b}), one between the states
$|\bx \rangle$, $|\bx' \rangle$, and the other one between the states $|{\tl \bx} \rangle$, $|{\tl \bx}' \rangle$,
which are all particular cases of a generic single particle state
\be
  |\psi (0) \rangle = \int \dd^3 \bx f(0,\bx) a^\dg (\bx) \vac = \int \dd^3 \bx\, 2 \om_\bx \tlf (0,\bx) \tla^\dg (\bx) \vac
\lbl{A9}
\ee
for two different choices, (\ref{A2}) and (\ref{A7}), of the wave packet profiles.

Explicit expression for $G(t,\bx':t,\bx)$ is given by the expression (\ref{5.7}), or the corresponding
three dimensional expression considered in Ref.\,\ci{Cirilo-Lombardo}, whilst the explicit expression
for the propagator (\ref{A8b}) is\,\ci{Padmanabhan,Ahluwalia}
\be
  {\tl G} (t',\bx';t,\bx) = \frac{1}{\pi^2} \frac{m^2}{\sqrt{r^2 - t^2}} K_1 \left ( m \sqrt{r^2 - t^2} \right ),
 ~~~ r^2 = (\bx' - \bx)^2.
\lbl{A10}
\ee

From the latter expression it follows that the amplitude for the transition between the events
separated by a space-like interval does not vanish. This fact has been explored within the
context of the Dirac field in Ref.\,\ci{Ahluwalia},
where it was argued that contrary to the common understanding conveyed in the modern literature,
such effect may have observable macroscopic consequences.

\vs{5mm}

\centerline{Acknowledgement}

This work has been supported by the Slovenian Research Agency.


\begin{thebibliography}{12}
\bi{NewtonWigner} T. Newton and E. Wigner, Localized states for elementary systems, 
{\it Rev. Mod. Phys.} {\bf 21}, 400--406 (1949).

\bi{Wightman} A. S. Wightman, On the Localizability of Quantum Mechanical Systems, {\it Rev. Mod. Phys.}
{\bf 34}, 845--872 (1962).

\bi{Kalnay} A.J. K\'alnay, Lorentz-invariant localization for elementary systems,
{\it Phys. Rev. D} {\bf 1},1092--1104 (1969).

\bi{Ruijgrok}  T. W. Ruijgrok, On Localization in Relativistic Quantum Mechanics,
in {\it Theoretical Physics Fin de Si\`ecle},
volume 539 of {\it Lecture Notes on Physics} (Springer-Verlag, Heidelberg, 2000).

\bi{Barat} N. Barat and J. C. Kimball, Localization and causality for a free particle,
{\it Phys. Lett. A} {\bf 308}, 110--115 (2003)

\bi{Mir-Kasimov} R. M. Mir-Kasimov, The Newton-Wigner State Localization and the Commutativity of the
Configuration Space, {\it Physics of Particle and Nuclei Letters}
{\bf 3}, 280--289 (2006).

\bi{Cirilo-Lombardo} D. J. Cirilo-Lombardo, Relativistic dynamics, Green function and
pseudidifferential operators,  Journal of Mathematical Physics {\bf 57}, 063503 (2016); doi: 10.1063/1.4953368,
arXiv:1610.03624 [hep-th].

\bi{Herrmann} L. O. Herrmann,  Localization in Relativistic Quantum Theories,  PhilSci Archive (2010),
http://philsci-archive.pitt.edu/5427/

\bi{Fleming1} G. N. Fleming, Covariant Position Operators, Spin, and Locality, {\it Phys. Rev.} {\bf 137}, B188 (1965)

\bi{Fleming2} G. N. Fleming, Lorentz Invariant State Reduction and Localization, in {\it Proceedings of the
Biennial Meeting of the Philosophy of Science Association}, Vol. 1988, Volume Two:
Symposia and Invited Papers (1988), pp. 112--126.


\bi{Monahan} A. H. Monahan and M. McMillan, Lorentz boost of the Newton-Wigner
position operator, {\it Phys. Rev. A} {\bf 56}, 2563--2566 (1997).

\bi{WightmanSchweber} A. S. Wightman and S. S. Schweber, Configuration Methods
in Relativistic Quantum Field Theory I, {\it Phys. Rev.} {\bf 98}, 812--837 (1955).

\bi{Manoukian} E. B. Manoukian, Rediscovering the Newton-Wigner Operator from a
Space-time Description of Quantum Field Theory, {\it Nuovo Cim. A} {\bf 103},
1495--1497) (1990).

\bi{Teller} Paul Teller, {\it An Interpretative Introduction to Quantum Field Theory},
(Univ. Press, Princeton, 1995).

\bi{Buscemi} F. Buscemi and G. Compagno, Causality and localization operators,
{\it Phys. Lett. A} {\bf 334}, 357--362 (2005).

\bi{Hegerfeldt} G. C. Hegerfeldt, Remarks on causality and particle localization,
{\it Phys. Rev. D} {\bf 10}, 3320 (1974).

\bi{Hegerfeldt2} G. C. Hegerfeldt and S. N. M. Ruijsenaars, Remarks on
Causality, localization, and spreading of wave packets, {\it Phys. Rev. D} {\bf 22}, 377--384 (1980).

\bi{Rosenstein} B. Rosenstein and M. Usher, Explicit illustration of causality violation:
Noncausal relativistic wave-packet evolution, {\it Phys. Rev. D} {[\bf 36}, 2381--2384 (1987). 

\bi{Mosley} S. N. Mosley and J. E. G. Farina, Causality and the scalar field energy,
{\it J. Phys. A: Math. Gen.} {\bf 23}, 3991--3996 (1990).

\bi{Wagner} R.E. Wagner, B.T. Shields, M.R. Ware, Q. Su, and R. Grobe,
Causality and relativistic localization in one-dimensional Hamiltonians,
{\it Phys. Rev. A} {\bf 83}, 062106(1--8) (2011).

\bi{Eckstein} M. Eckstein, Causal evolution of wave packets, 
{\it Phys.\ Rev.\ A } {\bf 95}, no. 3, 032106 (2017)
  doi:10.1103/PhysRevA.95.032106
arXiv:1610.00764 [quant-ph].

\bi{Ruijsenaars} S. N. M. Ruijsenaars, On Newton-Wigner Localization and Superluminal
Propagation Speeds, Annals of Phys. {\bf 137}, 33--43 (1981).

\bi{Al-Hashimi} M. H. Al-Hashimi and U. -J. Wiese, Minimal position-velocity uncertainty wave
packets in relativistic and non-relativistic quantum mechanics, {\it Ann. Phys.} {\bf 324}, 2599--2621 (2009).

\bibitem{Gavrilov} 
  S.~P.~Gavrilov and D.~M.~Gitman,
  Quantization of point - like particles and consistent relativistic quantum mechanics,'
  Int.\ J.\ Mod.\ Phys.\ A {\bf 15}, 4499 (2000)
  [hep-th/0003112].

\bi{Peskin} M. E. Peskin and D. V. Schroeder, {\it An Introduction to Quantum Field Theory},
(Perseus Books Publishing, Massachusetts, 1995).

\bi{Horwitz} B. Rosenstein and L. P. Horwitz, Probability current versus charge current of a
relativistic particle, {\it J. Phys. A: Math, Gen.} {\bf 18}, 2115--2121 (1985).

\bi{Fonda} L. Fonda and G. C. Ghirardi, {\it Symmetry Principles in Quantum Physic},
(Decker, New York, 1970).

\bi{Valente} G. Valente, Does the Reeh-Schlieder theorem violate relativistic causality?,
Studies in History and Philosophy in Modern Physics {\bf 48}, 147-155 (2014).

\bi{Haag} R. Haag, Local Quantum Physics, (Springer-Verlag, Berlin1996).

\bi{Reeh-Schlieder} H. Reeh and S. Schlieder, Bemerkungen zur Unit\"ar\"aquivalenz von Lorentzinvarianten Feldern,
Nuov. Cim. {\bf 22}, 1051--1068 (1961).



\bi{Schweber}  S. S. Schweber, {\it An Introduction to Relativistic Quantum Field Theory},
(Row, Peterson and Company, 1961).

\bi{Nikolic1} H. Nikoli\' c, The general-covariant and gauge-invariant theory of quantum particles in classical backgrounds, {\it Int. J. Mod. Phys. D} {\bf 12}, 407--477 (2003).

\bi{Nikolic2} H. Nikoli\'c, Probability in relativistic Bohmian mechanics of particles and strings, 
{\it Found. Phys.} {\bf 38}, 869--881 (2008), Appendix A.

\bi{Karpov1} E. Karpov, G. Ordonez, T. Petrosky, I. Prigogine, and G. Pronko,
Causality, Delocalization and Possitivity of Energy, {\it Phys. Rev. A} {\bf 62},012103 (2000).

\bi{Karpov2} I. Antoniou, E. Karpov, and G. Pronko, Non-Locality in Electrodynamics, {\it Found. Phys.}
{\bf 31},1641--1655 (2001).

\bi{PavsicBook} M. Pav\v si\v c, The Landscape of Theoretical Physics: A Global View;
From Point Particles to the Braneworld and Beyond, in Search of a Unifying Principle (Kluwer, 2001).

\bi{WikiPlanck} Wikipedia, Planck units,   https://en.wikipedia.org/wiki/Planck\_units

\bi{PavsicLettNuovCim} M. Pav\v si\v c, Towards Understanding Quantum Mechanics, General Relativity and the Tachyonic Causality Paradoxes, {\it Lett. Nuov. Cim.} {\bf 30}, 111--119 (19 81).

\bi{Deutsch1991} D. Deutsch, Quantum Mechanics near Closed Timelike Lines, {\it Phys. Rev. D} {\bf 44},
3197--3217 (1991).

\bi{Deutsch1} D. Deutsch, Scientific  American, March 1994, pp.  68--74.

\bi{Nimitz} G. Nimitz, Do Evanescent Modes Violate Relativistic Causality?, {Lect. Notes. Phys.}
{\bf 702}, 506--531 (2006)

\bi{Recami} M. Z. Rached, E. Recami and F. Fontana, Superluminal Localized Solutions to the Maxwell equations propagating through normal (non-evanescent) regions, {Annales de la Fondation Louis de Broglie} {\bf 26},
541--554 (2001).

\bi{Everett1} H. Everett, ``Relative State'' Formulation of Quantum Mechanics, Rev. Mod.
Pys., 454--462 (1957).

\bi{Everett2} H. Everett, On the Foundations of Quantum Mechanics, Thesis, Princeton University (1956), pp 1--140.

\bi{Everett3} H. Everett, Theory of the Universal Wavefunction,
 in ``The Many-Worlds Interpretation of Quantum Mechanics"  (Eds. B. S. DeWitt and N. Graham, Princeton Univ. Press, 1973), pp. 3--140

\bi{DeWitt1} B. S. DeWitt, Quantum Mechanics and Reality, Phys. Today, 155--165 (1970)

\bi{DeWitt2} B. S. DeWitt,  The Many-Universes Interpretation of Quantum Mechanics, in ``The Many-Worlds Interpretation of Quantum Mechanics"  (Eds. B. S. DeWitt and N. Graham, Princeton Univ. Press, 1973), pp. 167--218

\bi{Deutsch2} D. Deutsch, The Fabric of Reality (Penguin Press, London, 1997)

\bi{Zeh} H. Zeh, On the Interpretation of Measurement in Quantum Theory,
Found. Phys. {\bf 1}, 69--76 (1970).

\bi{Foldy} L. L. Foldy, Synthesis of Covariant Particle equations, Phys. Rev. {\bf 102}, 568--581 (1956).

\bi{Padmanabhan} T. Padmanabhan, {\it Quantum Field Theory}, (Springer, 2016).

\bi{Ahluwalia} S.P. Horwath, D. Schritt and D.V. Ahluwalia, Amplitudes for space-like separations
and causality, arXiv: 1110.1162 [hep-ph]


\end{thebibliography}
\end{document}